%% file: main.tex
\documentclass[letterpaper,twocolumn,10pt]{article}
\usepackage{usenix-2020-09}

\usepackage{tikz}
\usepackage{amsmath}
\usepackage{amssymb}
\usepackage{subfigure}
\usepackage{multirow}
\usepackage{ulem}

\begin{document}

\date{}

\title{\Large \bf ATP: Adaptive Tensor Parallelism for Foundation Models}

\author{
{\rm Shenggan Cheng}\\
School of Computing \\
National University of Singapore
\and
{\rm Ziming Liu}\\
School of Computing \\
National University of Singapore
\and
{\rm Jiangsu Du}\\
School of Computer Science and Engineering \\
Sun Yat-sen University
\and
{\rm Yang You}\\
School of Computing \\
National University of Singapore
} 


\maketitle

\input{tex/0_abstract.tex}

\input{tex/1_introduction.tex}
\input{tex/2_background.tex}
\input{tex/3_method.tex}
\input{tex/4_implementation.tex}
\input{tex/5_experiment.tex}
\input{tex/6_conclusion.tex}

\bibliographystyle{plain}
\bibliography{main}

\end{document}

%% file: tex/0_abstract.tex
\begin{abstract}
Foundation models have impressive performance and generalization capabilities across a wide range of applications. The increasing size of the models introduces great challenges for the training. Tensor parallelism is a critical technique that is currently used in almost all foundation model training and has a significant impact on overall training performance. However, current tensor parallelism in machine learning frameworks misses optimization opportunities in fitting various interconnection topologies. In this work, we present ATP, an adaptive tensor parallelism framework for foundation models, which can automatically select the optimal parallel strategy on different interconnections. We propose \textit{column- and row-first tensor parallelism} based on 2D device meshes and construct a search space. Combined with the \textit{hierarchical communication matrix}, ATP can identify the optimal strategy in the search space. We also propose chunk-based overlapping to reduce communication overhead. Our evaluations show ATP consistently outperforms the state-of-the-art approaches for various model sizes and interconnects, achieving end-to-end training performance improvements of up to 37-64\% on specific interconnects. Based on our theoretical model, the communication overhead of ATP decreases with scaling, indicating a qualitative leap forward.

\end{abstract}

%% file: tex/1_introduction.tex
\section{Introduction}


The term \textit{foundation models} \cite{Bommasani2021OnTO} refers to pretrained models with large-scale parameters that can be adapted to a wide range of downstream tasks. These models have impressive performance with generalization capabilities and have led to state-of-the-art results on benchmarks of natural language processing (BERT \cite{Devlin2019BERTPO}, GPT-3 \cite{brown2020language}) and computer vision (CLIP \cite{Radford2021LearningTV}, Florence \cite{Yuan2021FlorenceAN}). The most obvious trend in recent years is the rapid increase in the size of foundation models: from 100 billion \cite{brown2020language, Wu2021Yuan1L, Zhang2022OPTOP} to over 500 billion \cite{Smith2022UsingDA, Chowdhery2022PaLMSL}. 

Training foundation models, such as GPT-3 (which has 175 billion parameters), can be challenging due to the huge computational costs and memory capacity bottlenecks. For example, it would take about 288 years to train GPT-3 using a single NVIDIA V100, and the model's parameters would not fit in the main memory of even the most advanced GPUs (such as NVIDIA 80 GB A100). To address these issues, various parallelism techniques have been proposed, including tensor (intra-layer) parallelism \cite{Shazeer2018MeshTensorFlowDL, megatron}, pipeline (inter-layer) parallelism \cite{huang2019gpipe, narayanan2019pipedream} and ZeRO redundant optimizer \cite{Rajbhandari2019ZeROMO, Rajbhandari2021ZeROInfinityBT}. These approaches aim to efficiently parallelize computation and distribute parameters among multiple devices.

Among these, tensor parallelism is the most important training technique. This approach evenly distributes both computations and parameters across multiple devices. In the training of foundation models such as GPT-3 \cite{brown2020language}, OPT \cite{Zhang2022OPTOP}, Yuan \cite{Wu2021Yuan1L}, Megatron-Turing \cite{Smith2022UsingDA}, tensor parallelism and pipeline parallelism are often used in combination. Tensor parallelism splits the matrix multiplications across different devices and is typically used at the intra-node level, while pipeline parallelism is used at the inter-node level. Empirical results have shown that tensor parallelism should be used as much as possible without introducing cross-node communication, meaning that tensor parallelism should generally be used up to degree-$N$ when using $N$-GPU servers \cite{megatron}. In automatic parallelism approaches, tensor parallelism is a key design factor for the search space. For example, in Alpa \cite{Zheng2022AlpaAI}, the search space includes both inter-operator and intra-operator parallelism, where intra-operator parallelism represents tensor parallelism. Tensor parallelism is not only used for training, but also for inference to reduce latency and provide additional memory capacity across GPUs to fit parameters \cite{Pope2022EfficientlyST}.

However, existing approaches has several drawbacks. \textbf{1)} they only consider simple interconnect topologies, such as NVSwitch\cite{nvswitch}, and cannot adapt to more complicated topologies. For example, when performing tensor parallelism on multi-node systems, the all-reduce operation is limited by the slowest link in the topology, preventing it from exploiting the high-bandwidth NVLink within the server. \textbf{2)} tensor parallelism relies heavily on high-bandwidth interconnects, such as NVLink. On servers that are not equipped with NVLink, the communication costs introduced by tensor parallelism can become a significant bottleneck for training. \textbf{3)} unlike data parallelism, the communication introduced by tensor parallelism is synchronous, which makes it more difficult to overlap with computations, leading to more overhead. As a result, tensor parallelism is not as scalable as data parallelism.

To address these limitations, we present \textit{Adaptive Tensor Parallelism (ATP)} for foundation models. Unlike existing tensor parallelism approaches that have a fixed communication pattern, ATP can select different communication patterns on different topological interconnections. We propose column- and row-first tensor parallelism based on 2D device meshes. Different 2D device meshes form the search space, and we use the hierarchical communication matrix to estimate the communication cost of different strategies. This allows ATP to identify the best strategy in the search space, making it a topo-aware approach. In addition, we use chunk-based overlapping to reduce communication overhead. ATP not only outperforms state-of-the-art approaches but also has a simple API for users.

In summary, we make the following contributions:

\vspace{-\topsep}
\begin{itemize}
  \setlength{\parskip}{0pt}
  \setlength{\itemsep}{0pt plus 1pt}
    \item We propose column- and row-first tensor parallelism based on two-dimensional device meshes and construct a novel search space for tensor parallelism.
    \item We design \textit{hierarchical communication matrix} to describe the communication characteristics of complicated interconnects and identify the optimal strategy in the search space.
    \item We implement ATP with some communication optimizations, including chunk-based overlapping.
    \item ATP achieves up to 37-64\% improvement over the state-of-the-art approaches in specific interconnects. In theoretical analysis, the communication cost of ATP decreases with scaling on some topologies, representing a significant improvement.
\end{itemize}
\vspace{-\topsep}


%% file: tex/2_background.tex
\section{Understanding Tensor Parallelism}

\subsection{Tensor Parallelism on Transformer}
\label{2.1}

\begin{figure}[hbt]
    \centering
    \includegraphics[width=0.45\textwidth]{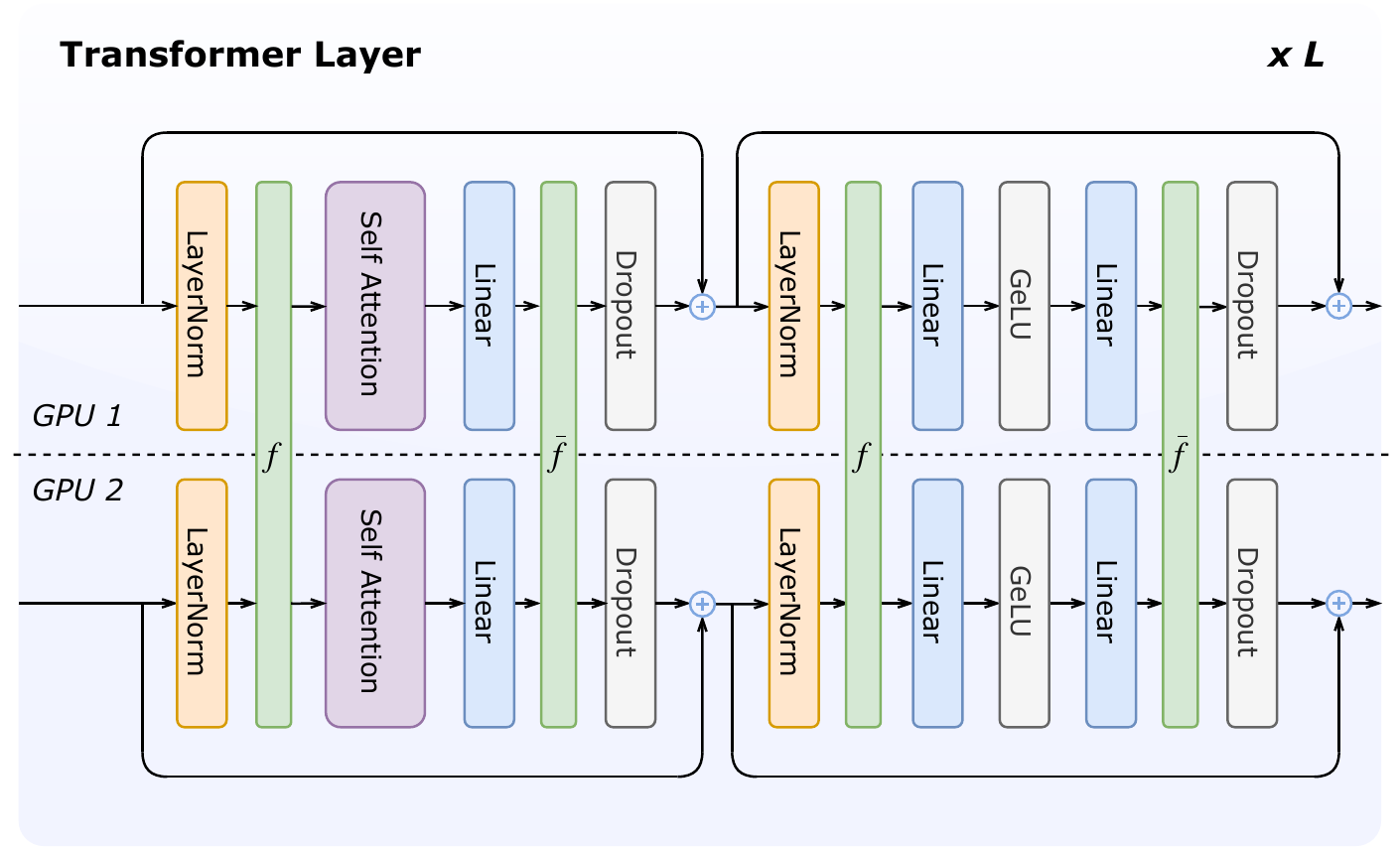}
    \caption{Transformer layer with tensor parallelism on two GPUs. $f$ and $\overline{f}$ are conjugate communication operators. In the forward pass, $\overline{f}$ represents an all-reduce operation, while in the backward pass $f$ represents an all-reduce operation.}
    \label{fig:transformer_arch}
\end{figure}

Transformer has become the bedrock of foundation models due to their excellent ability to model sequences. Taking a language model as an example, the input tokens are first fed into an embedding layer and then passed through a single-stack transformer encoder or decoder with the $L$ layers (see Figure \ref{fig:transformer_arch}). The input of transformer layer is a 3D tensor of size $[b, s, h]$ where $b, s, h$ are batch, sequence and hidden-size dimensions. Each transformer layer consists of a multi-head attention (MHA) block followed by a feed-forward block. In the MHA block, the sequence data is fed into three different MLP layers to obtain \textit{Query(Q)}, \textit{Key(K)}, \textit{Value(V)}. Then divided into $a$ heads, each with a hidden size of $d = h / a$. For each head, the attention score is calculated using Formula \ref{mha}. The feed-forward block has two layers of multi-layer perceptron (MLP). The first layer increases the hidden size to $4h$ and the second layer reduces it back to $h$.

\vspace{-5pt}
\begin{equation}\label{mha}
Att(Q, K, V) = softmax(\frac{QK^T}{\sqrt{d}})V \quad Q, K, V \in R^{s \times d}
\end{equation}

Tensor Parallelism, which is commonly used for training foundation models, was proposed by Megatron-LM \cite{megatron}. As shown in Figure \ref{fig:transformer_arch}, tensor parallelism parallelizes the MHA and feed-forward blocks. In the feed-forward block, there are two MLP layers and an activation function (GeLU):

\[Y={\rm GeLU}(XA), \quad Z=YB\]

where $A$ and $B$ are the weight of two MLP layers. We can split $A$ by columns and $B$ by rows and parallelize the computation into two parts:

\begin{align}
  A &= [A_1, A_2], \quad B = \begin{bmatrix} B_1 \\ B_2 \end{bmatrix} \nonumber\\
  [Y_1, Y_2] &= [{\rm GeLU}(XA_1), {\rm GeLU}(XA_2)] \nonumber\\
  Z &= {\rm reduce}(Y_1 B_1, Y_2, B_2) \nonumber
\end{align} 

MHA blocks can be parallelized in a similar way by dividing the weight matrix of $Q, K, V$ by columns and the output linear layer by rows.

This approach introduces \textit{two all-reduce} operations in both the forward and backward passes of each layer, distributing computation and memory footprints across devices. Inside the MHA and feed-forward blocks, tensor parallelism parallelizes model parameters, optimizer state, and activations. The LayerNorm, including the input/output activations, is duplicated in each tensor-parallel worker.

Several recent works, such as 2D tensor parallelism \cite{tp-2d} use the Scalable Universal Matrix Multiplication Algorithm (SUMMA) \cite{Geijn1995SUMMASU}. We regard 2.5D tensor parallelism as an extension of 2D tensor parallelism since it can be seen as a combination of 2D Tensor Parallelism and Data Parallelism. 2D Tensor Parallelism has two main issues: 1) broadcast of the weight matrix is expensive because the size of the weights is much larger than the activation of the giant model, and 2) multiple broadcasts in a single layer result in high overhead and low bandwidth utilization.

\subsection{Modern Accelerator Interconnect}

Due to the high demand on communication in distributed training, a number of accelerator-specific interconnect hardware has been developed and more complicated topologies have been introduced. Multi-GPU servers dominate the training hardware, so the interconnect architecture of GPUs is generally divided into two hierarchical levels, \textit{intra-node level} and \textit{inter-node level}.

\begin{figure}[hbt]
  \centering
  \subfigure[Server with NVSwitch] {
    \label{fig:intra-node-nvswitch}     
    \includegraphics[width=0.80\columnwidth]{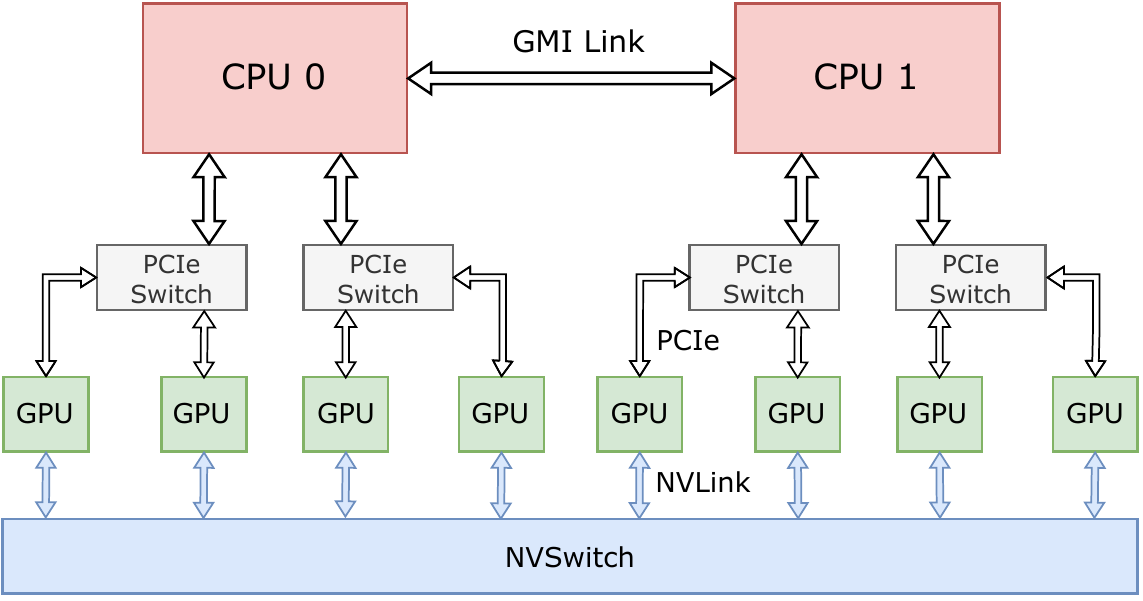}  
  }    
    
  \subfigure[Server with NVLink] { 
    \label{fig:intra-node-nvlink}     
    \includegraphics[width=0.80\columnwidth]{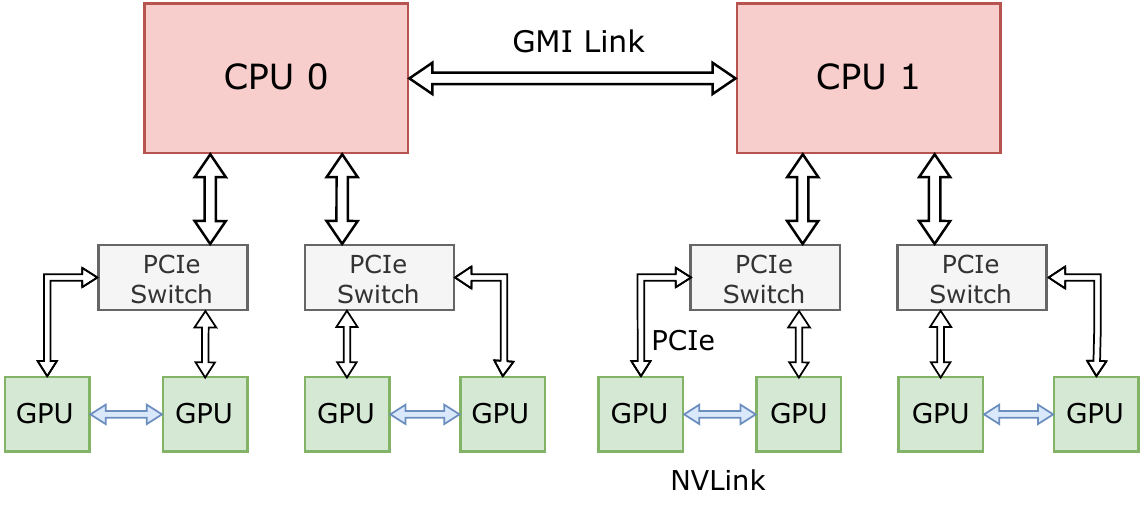}
  }    
  \caption{PCIe and NVLink topology for intra-node interconnect. (a) and (b) show the multi-gpu server with full interconnection and dual card interconnection.}     
  \label{fig:intra-node}
\end{figure}

\textbf{Intra-node Level.} At this hierarchy, training scales within a single node, where the interconnections between multiple GPUs are PCIe (64 GB/s for PCIe 4.0) or NVLink (600 GB/s for NVLink-v3). Figure \ref{fig:intra-node} shows the examples of these architectures. The PCIe network forms a balanced tree structure, where each GPU is connected to a PCIe switch, which is further connected to a CPU socket. The sockets are then bridged by GMI Link for $AMD$ $EPYC$ or QPI \cite{Ziakas2010IntelQI} for $Intel$ $XEON$. NVLink is a GPU-oriented interconnect technology proposed by NVIDIA with various configuration options. For example, in Figure \ref{fig:intra-node-nvswitch}, all GPUs are connected to the NVSwitch via NVLink, improving all-to-all communication capabilities. In Figure \ref{fig:intra-node-nvlink}, the eight GPUs are divided into four groups, with two GPUs in each group connected via NVLink. In this case, the GPU interconnect has a clear NUMA effect (closer proximity leads to stronger communication capabilities).

\begin{figure}[hbt]
  \centering
  \subfigure[Fat Tree] {
    \label{fig:inter-tree}     
    \includegraphics[width=0.30\columnwidth]{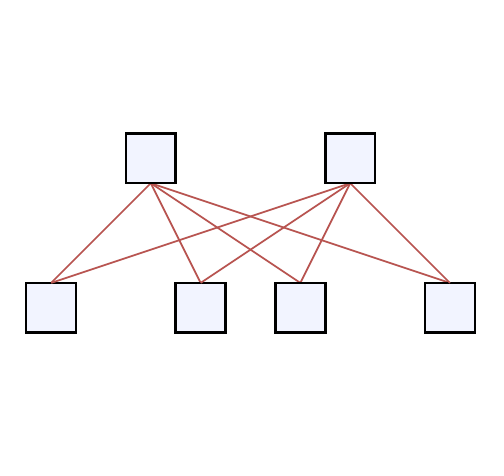}  
  }
  \subfigure[Torus] {
    \label{fig:inter-torus}     
    \includegraphics[width=0.30\columnwidth]{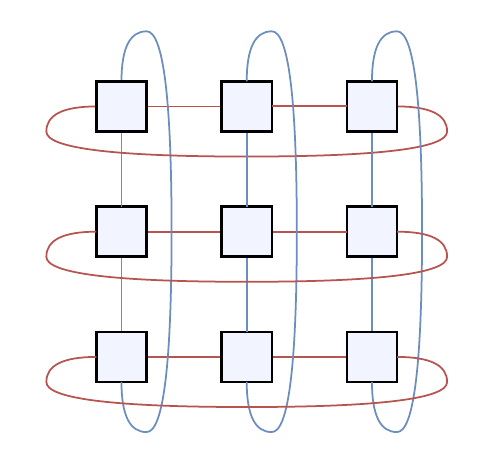}  
  }
  \subfigure[Dragonfly] {
    \label{fig:inter-dragonfly}     
    \includegraphics[width=0.30\columnwidth]{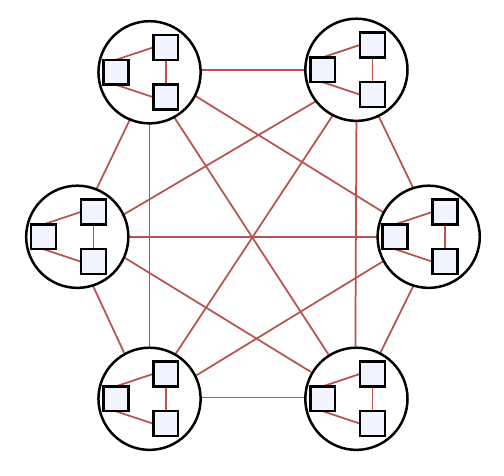}  
  }    
  \vspace{-5pt}
  \caption{Inter-node interconnect topology. }     
  \label{fig:inter-node}
\end{figure}

\textbf{Inter-node Level.} In data centers, the main interconnection technologies across nodes are Ethernet and InfiniBand \cite{InfiniBand2010IntroductionTI}. Cloud servers or supercomputers used for training are typically have at least 50 Gbps or 200 Gbps of communication bandwidth. There are many different configurations for topologies at this hierarchy, as shown in Figure \ref{fig:inter-node}, including fat tree, torus, and dragonfly. There are also many accelerator-specific direct interconnects at this level, such as the 2D Torus Inter-Core Interconnect (ICI) links in TPU pods that directly connects up to 1024 TPUv3 cores \cite{Jouppi2020ADS} and NVIDIA's upcoming Nvlink-Network Switch \cite{ishii2022nvlink}, which can also connect hundreds of GPUs directly via high-bandwidth NVLink \cite{ishii2022nvlink}. The development of new interconnected hardware and topologies presents opportunities and challenges for software-hardware co-design \cite{10.5555/3571885.3571899}.

\subsection{Communication Analysis}


\textbf{Communication Pattern.} Similar to data parallelism, tensor-parallel communication requires all-reduce operations. However, because the computations are interdependent with the communication, the all-reduce here must be synchronous. Tensor-parallel communication relies more on high-bandwidth communication because it cannot overlap with backward computation as data-parallel communication does.

\textbf{Communication Cost.} As mentioned in Section \ref{2.1}, the size of the tensor for each all-reduce is $[b, s, h]$, where $b,s,h$ are batch, sequence and hidden-size dimensions. In a model with $L$ layers, a total of $4L$ all-reduce operations are required in the forward and backward passes. Assuming an all-reduce bandwidth of $B$, the communication cost for one step is $4Lbsq/B$. In training foundation models, synchronous all-reduce of dozens of gigabytes of large tensors is required in one step, posing a high demand on interconnect bandwidth.

\textbf{All-reduce Performance.} Various vendors provide \textit{collective communications library (CCL)} to deliver intra-node and inter-node level communication capabilities for deep learning frameworks, such as NCCL on the NVIDIA platform. These high-performance all-reduce implementations are primarily based on the ring algorithm \cite{Sergeev2018HorovodFA}. The performance of ring all-reduce depends on the bandwidth of the slowest link on the ring, which may result in some local high-bandwidth interconnects being wasted. For example, in cross-server all-reduce communication, the all-reduce performance is limited by the bandwidth of the cross-server interconnect, and the high bandwidth of the NVLink inside the server is wasted.

%% file: tex/3_method.tex
\section{Adaptive Tensor Parallelism}

\subsection{Sharding Notion}

Many research and machine learning frameworks that use the concept of \textit{sharding} to describe how tensors are distributed across multiple devices in a parallel strategy, particularly in the context of auto-parallelism. In this paper, we will be focusing on the sharding notion from \textit{PyTorch Distributed Tensor}. 

\begin{figure}[hbt]
    \centering
    \includegraphics[width=0.40\textwidth]{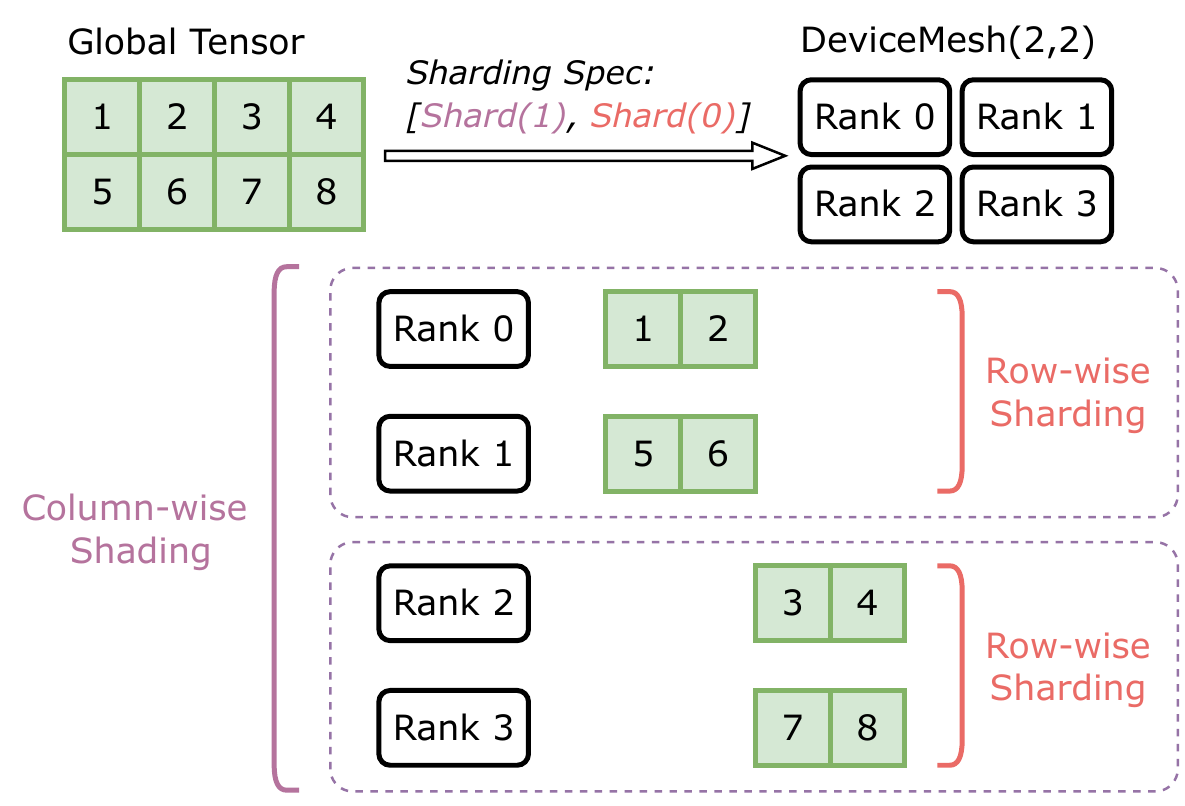}
    \vspace{-5pt}
    \caption{Sharding a 2D tensor on a 2D device mesh.}
    \label{fig:sharding_notion}
\end{figure}

\textbf{Device Mesh.} In the training of foundation models, it is often necessary to use a combination of different parallelization strategies, which involves communication between certain ranks. To manage these communication groups, we group the devices into multidimensional device meshes. A device mesh with $N$ devices can be expressed as $DeviceMesh(d_1, d_2, ... d_n)$. A device mesh can be thought of as having $n$ levels, with the devices in the current group being divided into $d_i$ sub-groups at the $i$-th level. The total number of devices in the mesh is equal to $N=d_1 \times d_2 \times .. \times d_n$. For example, a device mesh with four devices can be represented in four different ways: 1D: [4], 2D: [1,4], [2,2], [4,1]. $DeviceMesh(d_1, 4 / d_1)$ means that the four devices are divided into $d_1$ groups, with each group having $4/d_1$ devices for sharding. Figure \ref{fig:sharding_notion} shows a device mesh with $d_1 = 2$.

\textbf{Sharding Spec.} We use the sharding spec to describe the strategy for distributing the global tensor across the devices in a device mesh. Like Alpa \cite{Zheng2022AlpaAI} and OneFlow \cite{Yuan2021OneFlowRT}, we have chosen to use three types of \textit{placement strategies}: \textit{Shard}, \textit{Replicate}, and \textit{Partial}:

\vspace{-\topsep}
\begin{itemize}
    \setlength{\parskip}{0pt}
    \setlength{\itemsep}{0pt plus 1pt}
    \item \textit{Shard($d$)}: split the tensor along the $d$-th dimension across devices.
    \item \textit{Replicate}: replicate the tensor across devices.
    \item \textit{Partial($op$)}: partial the tensor across devices, requiring all-reduce communication to obtain the global tensor. $Op$ represents the \textit{ReduceOp}, such as $SUM$ or $MAX$.
\end{itemize}
\vspace{-\topsep}

From the perspective of a device mesh, the sharding of a tensor can be thought of as the selection of a placement strategy at each level of the device mesh. Therefore, the sharding spec can be viewed as a sequence of placement strategies, the length of which is equal to the number of dimensions of the device mesh. For an $N$-dimensional device mesh, the sharding spec of the global tensor can be defined as $[P_1, P_2, ... P_N]$, where $P_i \in \{Shard(d), Replicate, Partial(op)\}$ represents the placement strategy for the $i$-th dimension of the device mesh. Figure \ref{fig:sharding_notion} illustrates a valid sharding spec for a 2D tensor on a 2D device mesh. In the sharding spec $[Shard(1), Shard(0)]$, $Shard(1)$ means that the tensor is column-wise partitioned into two parts for rank-[0,1] and rank-[2,3], and then $Shard(0)$ means that each of these partitioned tensors is further row-wise partitioned for each pair of ranks. If the sharding spec is $[Replicate, Shard(0)]$, the sharded tensor for rank-0 and rank-2 will be $[1,2,3,4]$ and rank-1 and rank-3 will be $[5,6,7,8]$. Unlike the sharding spec in Alpa, which binds the placement strategy to the tensor dimensions, we have chosen to bind the strategy to the dimensions of the device mesh. This approach is more concise and easier to understand.

The parallelism strategy can be represented as a sharding spec on a 1D device mesh. Table \ref{tab:1d_sharding} illustrates the sharding specs of the input, weight, and output in data parallelism (DP) and tensor parallelism in Megatron-LM (Column TP and Row TP). For the combination of DP and TP, a 2D device mesh is typically used, with the first dimension of the sharding spec corresponding to DP (as shown in the first row of the table) and the second dimension corresponding to TP (as shown in the second or third row of the table).

\vspace{-5pt}
\begin{table}[hbt]
\small
\centering
\label{tab:1d_sharding}
\caption{Sharding specs for MLP layer on 1D device mesh.}
\renewcommand\arraystretch{1.2}
\begin{tabular}{rlll}
\noalign{\hrule height 1pt}
Parallelism   & Input           & Weight          & Output             \\ \hline
DP            & $[Split(0)]$    & $[Replicate]$   & $[Split(0)]$       \\
Column TP     & $[Replicate]$   & $[Split(1)]$    & $[Split(1)]$       \\
Row TP        & $[Split(1)]$    & $[Split(0)]$    & $[Partial(SUM)]$   \\ 
\noalign{\hrule height 1pt}
\end{tabular}
\end{table}

\subsection{Search Space for Tensor Parallelism}
\label{3.2}

\begin{figure*}[hbt]
    \centering
    \includegraphics[width=0.92\textwidth]{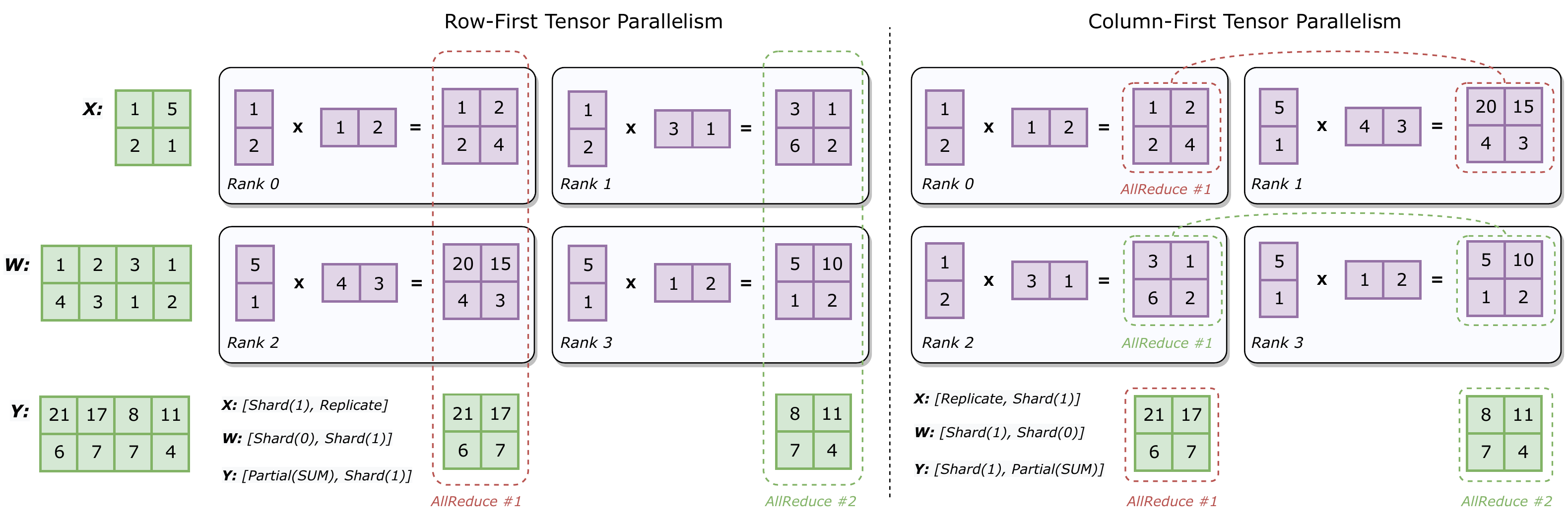}
    \caption{Row-First and Column-First Tensor Parallelism for $Y=XW$ on $DeviceMesh(2,2)$. The green matrix represents the global tensor and the purple matrix represents sharded tensor. These two approaches have different priorities in the sharding of the weight matrix ($W$). A grouped all-reduce communication is required to obtain the global tensor of $Y$.}
    \label{fig:atp-2d}
\end{figure*}

With the definition of device mesh and sharding spec, we can express more complex parallelization strategies. In this section, we will introduce two types of tensor parallelism algorithms on 2D device meshes, and demonstrate how they can be applied to transformer models. Finally, we will discuss the search space for tensor parallelism.

In Megatron-LM's tensor parallelism, the weight matrix ($W$) is split row-wise or column-wise on a 1D device mesh. When using a 2D device mesh, $W$ is split at two levels. As shown in Figure \ref{fig:atp-2d}, there are two sharding specs for $W$, $[Shard(0), Shard(1)]$ and $[Shard(1), Shard(0)]$. In the first spec, the matrix is split row-wise at the first level of the device mesh, and then column-wise at the second level. The other spec splits column-wise at the first level and then row-wise at the second level.

Then we consider how to perform the distributed GEMM $Y=XW$ on $DeviceMesh(d_1, d_2)$. Based on the two sharding specs for the weight matrix, two tensor parallelism approaches can be proposed: \textit{row-first tensor parallelism} and \textit{column-first tensor parallelism}. As shown in the left half of Figure \ref{fig:atp-2d}, in row-first tensor parallelism, the sharding specs of $W$ and $X$ are $[Shard(0), Shard(1)]$ and $[Shard(1), Replicate]$. The local GEMM operation at each rank results in a sharded $Y$ with the sharding spec $[Partial(SUM), Shard(1)]$. To obtain the global tensor result of $Y$, it is necessary to perform an all-reduce operation on the first dimension of the device mesh and then an all-gather operation on the second dimension. Column-first tensor parallelism uses a different sharding strategy for the weight matrix, and the dimensions of the device mesh are transposed in the communication, resulting in a different communication pattern compared to row-first tensor parallelism.

For row-first tensor parallelism, when the size of the global tensor $X$ is $[b, h_1]$, the size of $W$ is $[h_1, h_2]$, and the size of $Y$ is $[b, h_2]$, the sharded $X$ has a size of $[b, h_1 / d_1]$, the sharded $W$ has a size of $[h_1 / d_1, h_2 / d_2]$, and the sharded $Y$ has a size of $[b, h_2 / d_2]$. The GEMM on global tensors requires $b \times h_1 \times h_2$ fused multiply-add (FMA) operations, while each rank requires $b \times h_1 \times h_2 / (d_1 \times d_2)$ FMA operations. The computations are evenly distributed across the ranks, requiring communication on a tensor of size $[b, h_2 / d_2]$. In contrast, for column-first tensor parallelism, the size of the tensor for communication is $[b, h_2 / d_1]$.

\subsubsection{Transformer}

As introduced in Section \ref{2.1}, the transformer layer consists of an attention block followed by a feed-forward block. We will describe how to use row- and column-first tensor parallelism in these two blocks.

\begin{figure}[hbt]
  \centering
  \subfigure[Attention block] { 
    \label{fig:atp-attention}     
    \includegraphics[width=0.84\columnwidth]{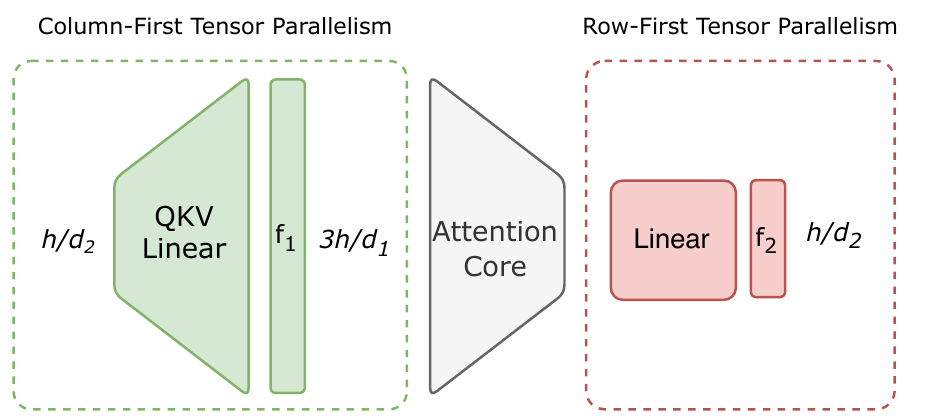}
  }
  \subfigure[Feed-forward] {
    \label{fig:atp-mlp}     
    \includegraphics[width=0.75\columnwidth]{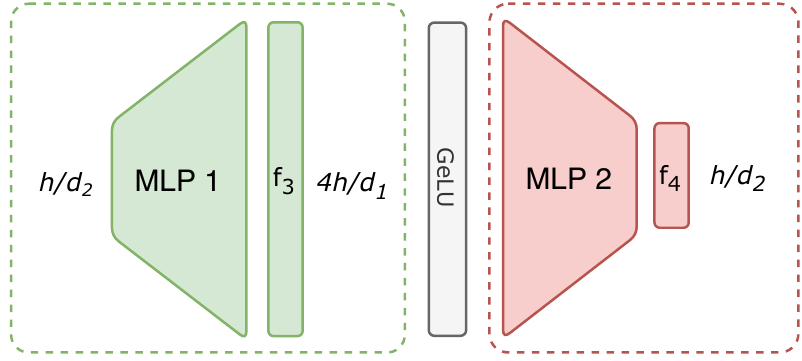}  
  }    
  \caption{Parallel transformer layer with row- and column-first tensor parallelism. $f_{1,3}$ and $f_{2,4}$ are conjugate operations: $f_{1,3}$ performs an all-reduce on the first dimension of the device mesh in the forward pass and on the second dimension in the backward pass, while $f_{2,4}$ is the reverse.}     
  \label{fig:atp-transformer}
\end{figure}

\textbf{Feed-forward.} There are two MLP layers and an activation function in feed-forward ($Y = {\rm GeLU}(XA), Z = Y B$). Megatron-LM uses column parallelism for the first MLP layer and row parallelism for the second. Similarly, in our approach, we use column-first tensor parallelism for the first MLP layer and row-first for the second, as shown in Figure \ref{fig:atp-mlp}. The sharded feed-forward block can be expressed as follows:

\vspace{-8pt}
\begin{align}
  Y_{shard} &= f_3(X_{shard}A_{shard}) \nonumber\\
  Z_{shard} &= f_4({\rm GeLU}(Y_{shard})B_{shard}) \nonumber 
\end{align} 

The sharding specs of weight matrix $A_{shard}$ and $B_{shard}$ are $[Shard(1), Shard(0)]$ and $[Shard(0), Shard(1)]$, respectively. The input($X_{shard}$) and output tensor($Z_{shard}$) have the same sharding spec: $[Replicate, Shard(1)]$. $f_3$ and $f_4$ are two conjugate communication operations: $f_3$ performs an all-reduce on the first dimension of the device mesh in the forward pass and on the second dimension in the backward pass, while $f_4$ is the reverse. 

\textbf{Attention block.} In this block, we use a similar strategy to the feed-forward block, using column-first tensor parallelism in the \textit{QKV Linear} and row-first in the \textit{Output Linear}, as shown in Figure \ref{fig:atp-attention}. The main difference is that we want the computation of the \textit{Attention Core} to be fully sharded, so the sharding spec of the input tensor should not include $Replicate$. To achieve this, we scatter the input tensor before the \textit{Attention Core}, and gather the output tensor before the \textit{Output Linear}. $f_1$ and $f_2$ are two conjugate communication operations, similar to $f_3$ and $f_4$ in the feed-forward block. $f_1$ has a smaller amount of communication compared to $f_3$ because the output hidden size of the \textit{QKV Linear} is smaller.

Different device meshes correspond to different tensor sizes and communication groups for row- and column-first tensor parallelism, which can result in different optimal tensor parallelism strategies for different interconnect topologies. For example, if we have $2^n$ devices, then we will have $n+1$ kinds of 2D device meshes. This forms a search space in which we need to find the device mesh with the optimal performance for tensor parallelism in a specific topology.

Under the $DeviceMesh(N, 1)$ device mesh, our approach is similar to the tensor parallelism in Megatron-LM \cite{megatron}. Google proposed a novel tensor parallelism algorithm \cite{Pope2022EfficientlyST} to reduce the inference latency of the giant transformer model on TPU clusters. On $q^2$ TPUs with a 2D torus topology, Google's algorithm is similar to our proposed tensor parallelism on $DeviceMesh(q / 2, 2q)$. These tensor parallelism methods are all contained in our search space, which demonstrates the potency of our search space and guarantees that the final performance will not be worse than these traditional tensor parallelism methods.

\subsection{Theoretical Analysis}
\label{3.3}

To analyze the communication cost theoretically, we assume that $B_1$ is the all-reduce bandwidth on the first dimension of the device mesh and $B_2$ is the bandwidth on the second dimension. For an $L$-layer transformer model with an input size of $[b, s, h]$, where $b$, $s$, and $h$ are the batch, sequence, and hidden size dimensions, the communication time of the all-reduce introduced by tensor parallelism is:

\vspace{-8pt}
\begin{align}
  T_{comm} &= 2Lbs \times (T_{f_1} + T_{f_2} + T_{f_3} + T_{f_4}) \nonumber\\
  &= 2Lbs \times (\frac{3h}{d_1B_2} + \frac{h}{d_2B_1} + \frac{4h}{d_1B_2} + \frac{h}{d_2B_1}) \nonumber \\
  &= 2Lbs \times (\frac{7h}{d_1B_2} + \frac{2h}{d_2B_1}) \label{tcomm}
\end{align} 

It is worth noting that as the number of tensor-parallel workers ($d_1 \times d_2$) increases, the communication time ($T_{comm}$) decreases. This communication characteristic helps to extend tensor parallelism to a larger scale.

From the perspective of the communication group, Megatron-LM requires the participation of all tensor-parallel workers. On a 2D device mesh, communication is grouped, meaning that each communication requires only one row or column of workers on the device mesh. This communication pattern has the potential to better exploit complicated interconnect topologies.

\subsection{Hierarchical Communication Matrix}
\label{3.4}

As previously mentioned, the goal is to find the optimal device mesh that provides the best performance in a specific topology. Currently, many performance analysis and automatic parallelism frameworks for model training rely on communication volume to measure communication cost. However, to more effectively find the optimal device mesh, we utilize \textit{hierarchical communication matrix} to account for communication costs in complex topologies.

\begin{figure}[hbt]
  \centering
  \subfigure[Interconnect Topology 1] { 
    \label{fig:hcm-1}     
    \includegraphics[width=0.72\columnwidth]{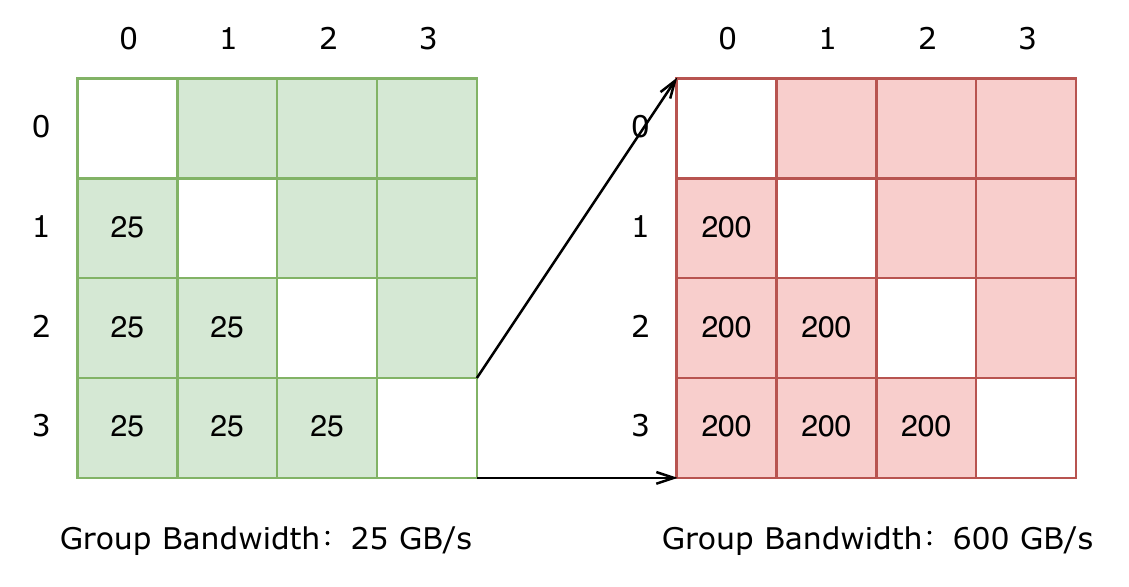}
  }
  \subfigure[Interconnect Topology 2] {
    \label{fig:hcm-2}     
    \includegraphics[width=0.72\columnwidth]{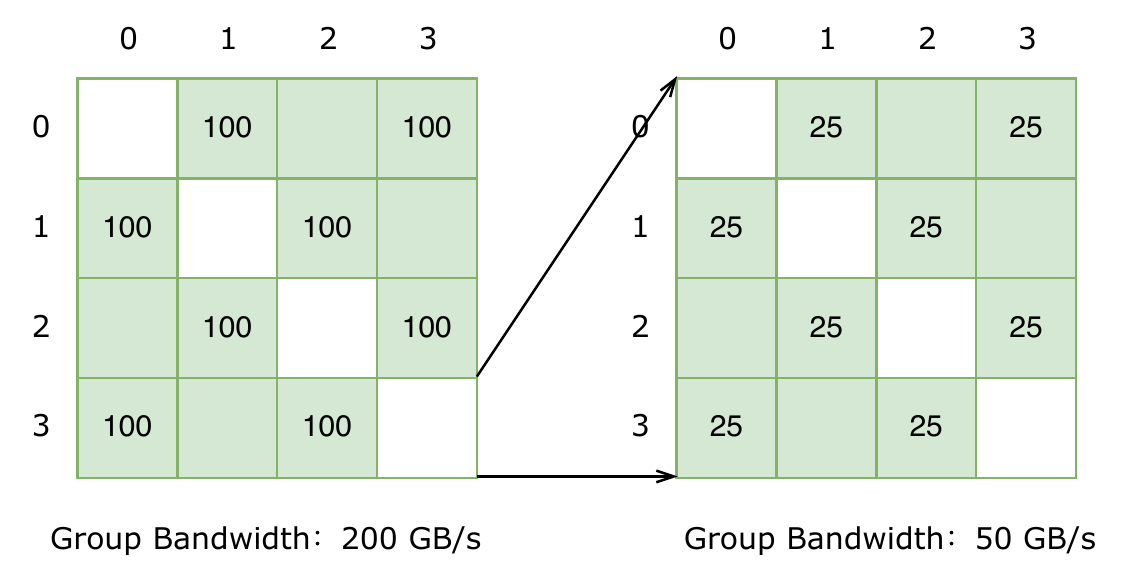}  
  }    
  \caption{Examples of hierarchical communication matrix. (a) is a cluster of four nodes connected via 200 Gbps HDR, with four GPUs per node interconnected using NVLinkv3. (b) is $4 \times 4$ devices with 2D Torus interconnect.}
  \label{fig:hcm-example}
\end{figure}

One of the main features of modern accelerator interconnects is their hierarchical nature. Within nodes, multiple GPUs are connected using PCIe or NVLink, and different nodes are connected using InfiniBand or Ethernet. The traditional method of describing the communication capabilities of a topology is through a P2P communication matrix, which is a $N \times N$ matrix that represents the bandwidth of the interconnections between the $N$ ranks. However, we have introduced the concepts of \textit{hierarchy} and \textit{group bandwidth} into the communication matrix to more accurately represent the communication capabilities of the topology. Figure \ref{fig:hcm-example} shows some examples of hierarchical communication matrix.

The hierarchical communication matrix has multiple layers, each containing a P2P communication matrix and group bandwidth. At higher layers, each rank may represent a group of devices. For example, in Figure \ref{fig:hcm-1}, there are two layers, with each rank representing a node (consisting of 4 devices) at the higher level and a single device at the lower level. The P2P communication matrix is the aggregate bandwidth between the two ranks, which at a higher level is the two groups of devices. The group bandwidth for each rank represents the aggregate bandwidth of that group of devices to the outside world. In Figure \ref{fig:hcm-1}, we can see that there are 4 GPUs interconnected by 4 NVLinks, each providing 50 GB/s of bandwidth. Therefore, the P2P communication bandwidth is 200 GB/s, while the group bandwidth for each GPU is 600 GB/s. Figure \ref{fig:hcm-2} shows a 4x4 device mesh with a 2D Torus interconnect. Assuming that the bandwidth of each link is 25 GB/s, the second level of the matrix shows a torus ring in which devices $i$ and $i+1$, $i-1$ are connected. For the first level, there are 4 links between rings (one for each device), resulting in a P2P bandwidth of 100 GB/s. The links in both directions on the ring make the group bandwidth twice as large as P2P.

\subsection{Adaptive Tensor Parallelism}

By combining the search space (described in Section \ref{3.2}) and the hierarchical communication matrix (described in Section \ref{3.4}), we can propose Adaptive Tensor Parallelism (ATP) for automatically selecting the optimal tensor parallelism strategy. In the search space, we can estimate the communication time for each device mesh and identify the optimal mesh with the minimum communication time.

First, we estimate the bandwidth of all-reduce communication for each device mesh. As discussed in Section \ref{3.3}, on the device mesh, all-reduce is required for one column or one row of workers, whose bandwidths correspond to $B_{1}$ and $B_{2}$, respectively. We use the hierarchical communication matrix to estimate the bandwidth that can be provided by the interconnect link under all-reduce, resulting in $B_{1}^{'}$ and $B_{2}^{'}$, respectively. Assuming that the hierarchical communication matrix has $l$ layers with $R_i$ ranks per layer, and denoting the group bandwidth of layer $j$ as $GroupBW_j$, we can compute $B_1'$ and $B_2'$ as follows: when considering $DeviceMesh(d_1, d_2)$, the first dimension involves layers $1$ to $i$, and the second dimension is $i(+1)$ to $l$ ($i+1$ when $d_1 \neq \prod_{j: 1 \rightarrow i} R_j$). The performance of all-reduce depends on the bandwidth of the bottleneck, so $B_1'$ and $B_2'$ can be calculated using the following formula:

\vspace{-5pt}
\begin{align}
  B_1' &= min(GroupBW_{1 \rightarrow i} / d_2) \nonumber\\
  B_2' &= min(GroupBW_{i+(1) \rightarrow l}) \label{bwd}
\end{align} 

When performing all-reduce in the first dimension of the device mesh, there are $d_2$ groups of all-reduce, each with $d_1$ devices. Since $d_2$ groups of all-reduce share the interconnection bandwidth, so we need to divide the bandwidth by $d_2$ when calculating $B_{1}^{'}$. The P2P communication matrix can be used to correct the group bandwidth when the all-reduce algorithm cannot utilize the full group bandwidth. For example, considering $DeviceMesh(8, 2)$ in Figure \ref{fig:hcm-1}, $B_{2}^{'}$ corresponds to the bandwidth between the two GPUs, which is equal to 200 GB/s. Note that $B_{2}^{'}$ is lower than the group bandwidth (600 GB/s) due to the constraints of the P2P communication matrix. $B_{2}^{'}$ corresponds to 8-GPU inter-node communication, and since the two groups of all-reduce will share the inter-node bandwidth, $B_{1}^{'}$ is equal to 12.5 GB/s.

We assume that the cost of all-reduce follows the cost model of Rabenseifner’s algorithm \cite{Thakur2005OptimizationOC}. Ignoring the latency part, we can calculate the algorithm bandwidth $B_1$ and $B_2$ for $DeviceMesh(d_1, d_2)$ using the following formula:

\vspace{-10pt}
\begin{align}
B_1 = \frac{d_1}{2(d_1 - 1)} B_{1}^{'} \quad B_2 = \frac{d_2}{2(d_2 - 1)} B_{2}^{'} \label{allreduce}
\end{align} 

After obtaining $B_1$, $B_2$ from the hierarchical communication matrix, we can calculate the tensor parallel communication time based on Equation \ref{tcomm}. ATP selects the device mesh with the smallest $T_{comm}$ out of all the possible device meshes.

%% file: tex/4_implementation.tex
\section{Implementation}

\subsection{Chunk-based Overlapping}

Unlike data parallelism, which introduces all-reduce communication that can overlap with backward computation, tensor parallelism introduces synchronous all-reduce communication that can cause more severe bottlenecks. To reduce overhead, we propose a chunk-based overlapping technique in which the tensor is computed sequentially in chunks, creating overlapping opportunities


\begin{figure}[hbt]
    \centering
    \includegraphics[width=0.44\textwidth]{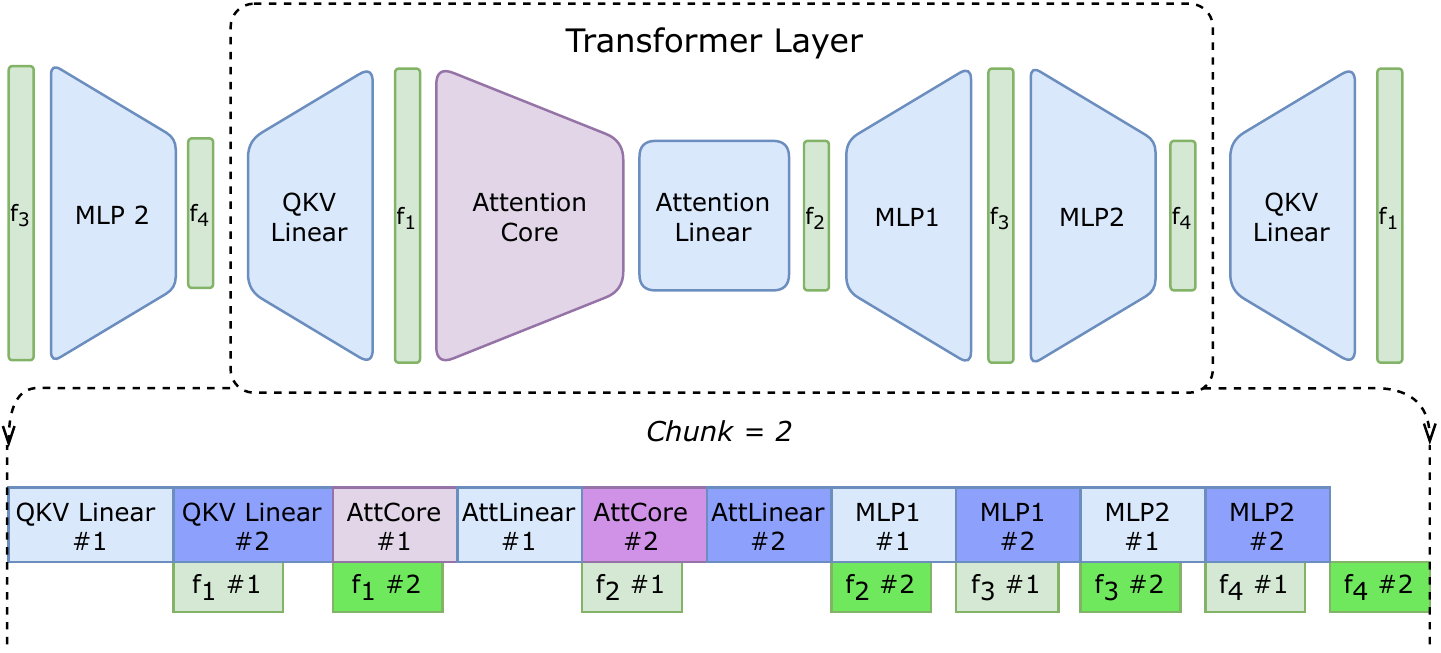}
    \caption{Chunk-based overlapping for one transformer layer. This example shows chunk sizes of 2. The lighter color is for computation and communication of the first chunk, and the darker color is for the second.}
    \label{fig:overlap}
\end{figure}

Since there is no dependency between the computation and communication of different samples in a batch, we can chunk on the batch dimension to reduce overhead. Figure \ref{fig:overlap} shows a chunk size of 2. For each block, the data from two chunks can be processed independently, with communication triggered after the first chunk is processed and the second chunk is processed simultaneously. While a larger chunk size allows for more overlapping, it can also lead to inefficient computation and communication. Therefore, in practice, it is usually sufficient to set the chunk size to 2 or 4.




Unlike some previous work that fused all-reduce with GEMM \cite{Jangda2021BreakingTC}, chunk-based overlapping combines the transformer model structure with overlapping on a larger scope.

\subsection{Other Details}

For linear layers, the chunk-based technique is used to overlap computation and communication in the forward pass. However, in the backward pass, the gradient of the input and the weight are two matrix products that do not depend on each other. The back-propagation equations for the matrix-matrix product $Y=XW$ are as follows:

\[\frac{\partial L}{\partial X} = \frac{\partial L}{\partial Y}W^T , \quad \frac{\partial L}{\partial W} = X^T\frac{\partial L}{\partial Y} \]

After computing the gradient of $X$, an all-reduce communication is required, which can be overlapped with the computation of the gradient of $W$. To prevent the communication kernel from being delayed and reducing the effectiveness of overlapping, we set the environment variable \texttt{CUDA\_DEVICE\_MAX\_CONNECTIONS} to 1.

\begin{figure}[hbt]
    \centering
    \includegraphics[width=0.42\textwidth]{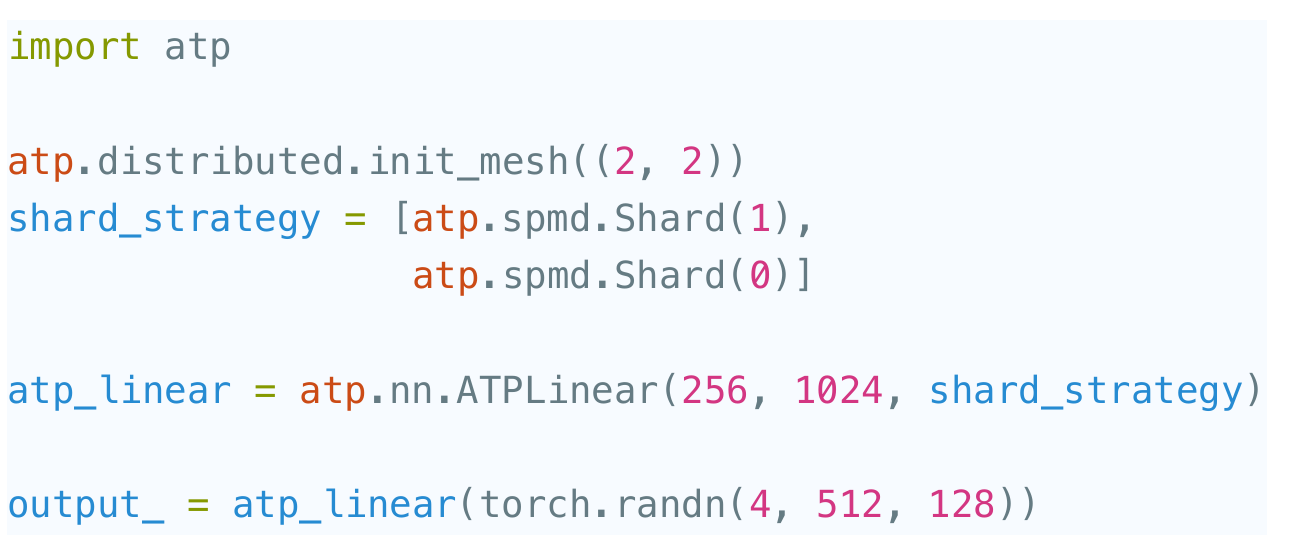}
    \caption{An example code snippet for ATP's API. In this example, the user creates a \textit{DeviceMesh(2,2)} using the \texttt{init\_mesh} and specifies the column-first tensor parallelism \texttt{ATPLinear}.}
    \label{fig:user-iterface}
\end{figure}

To reduce the development cost of distributed models, ATP offers easy-to-use APIs for higher-level frameworks. Figure \ref{fig:user-iterface} shows an example code of column-first tensor parallelism on $DeviceMesh(2,2)$. To use ATP, users simply call the \texttt{init\_mesh} function and use the \texttt{ATPLinear} module instead of the \texttt{Linear} module in PyTorch. The communication groups in the device mesh and the placement strategy are based on the PyTorch SPMD implementation (\url{https://github.com/pytorch/tau/tree/89700fd}).

%% file: tex/5_experiment.tex
\section{Evaluation}

The evaluation aim to answer the following questions:

\vspace{-\topsep}
\begin{itemize}
  \setlength{\parskip}{0pt}
  \setlength{\itemsep}{0pt plus 1pt}
  \item \textbf{Q1}: How does ATP compare with the state-of-the-art tensor parallelism approaches?
  \item \textbf{Q2}: What is the impact of the overlapping optimization?
  \item \textbf{Q3}: Does hierarchical communication matrix help ATP to select to the optimal strategy?
  \item \textbf{Q4}: How effective is ATP on different interconnect topologies, and how should future interconnect topologies be designed?
\end{itemize}
\vspace{-\topsep}

In our experiments, we use three different hardware environments: two single-node machines (\textit{Machine A} and \textit{Machine B}) and a multi-node cluster (\textit{Cluster C}). Both \textit{Machine A} and \textit{Machine B} are equipped with 8 NVIDIA A100 80 GB GPUs. \textit{Machine A} is equipped with NVSwitch and \textit{Machine B} is equipped with dual-GPU NVLink. The interconnection topology of these machines can be seen in Figures \ref{fig:intra-node-nvswitch} and \ref{fig:intra-node-nvlink}. \textit{Cluster C}, which is a multi-node cluster comprised of NVIDIA A100 GPUs, has a inter-node bandwidth of 200 Gbps.

In these three hardware environments, we conduct experiments using four different interconnects: (1) Machine A without NVLink ($\bf IC_1$), (2) Machine B with NVLink ($\bf IC_2$), (3) Machine A with NVLink ($\bf IC_3$), and (4) Cluster C with InfiniBand ($\bf IC_4$). We evaluate the performance of PCIe on Machine A by setting the NCCL environment variable \texttt{NCCL\_P2P\_DISABLE} to 1, which disables NVLink.

We use appropriately sized GPT models with half-precision (FP16). There are four sizes of GPT models ($\bf M_1$, $\bf M_2$, $\bf M_3$, $\bf M_4$) as listed in Table \ref{tab:model_list}. We use the metric of \textit{achieved teraFlOP/s per GPU} to evaluate performance. To calculate the number of floating-point operations (FLOPs) in a transformer layer, we use the formula from Megatron-LM\cite{megatron}. We consider FLOPs in both the forward and backward passes, but do not consider activation checkpointing and therefore do not need to multiply the forward FLOPs by 2 as in Megatron-LM. By default, we set the batch size to 4 and the sequence length to 2048.

\begin{table}[thb]
\small
\centering
\label{tab:model_list}
\caption{Four sizes of GPT models for evaluation.}
\renewcommand\arraystretch{1.1}
\begin{tabular}{lllll}
\noalign{\hrule height 1pt}
        & hidden size & heads & \begin{tabular}[c]{@{}l@{}}\#billion params \\ per layer\end{tabular} & \begin{tabular}[c]{@{}l@{}}\#TFLOPs \\ per layer\end{tabular} \\ \hline
\textbf{$\bf M_1$} & 2048        & 16    &  0.048                                                             &   2.625                                                           \\
\textbf{$\bf M_2$} & 4096        & 32    &  0.192                                                            &   9.75                                                           \\
\textbf{$\bf M_3$} & 8192        & 64    &  0.768                                                            &  37.5                                                            \\
\textbf{$\bf M_4$} & 12288       & 96    &  1.728                                                             &  83.25                                                            \\ \noalign{\hrule height 1pt}
\end{tabular}
\end{table}

\subsection{Comparison with the SOTA Methods}

\begin{figure*}[thb]
  \centering
  \subfigure[$\bf IC_1$] { 
    \label{fig:sota-t1}     
    \includegraphics[width=0.92\columnwidth]{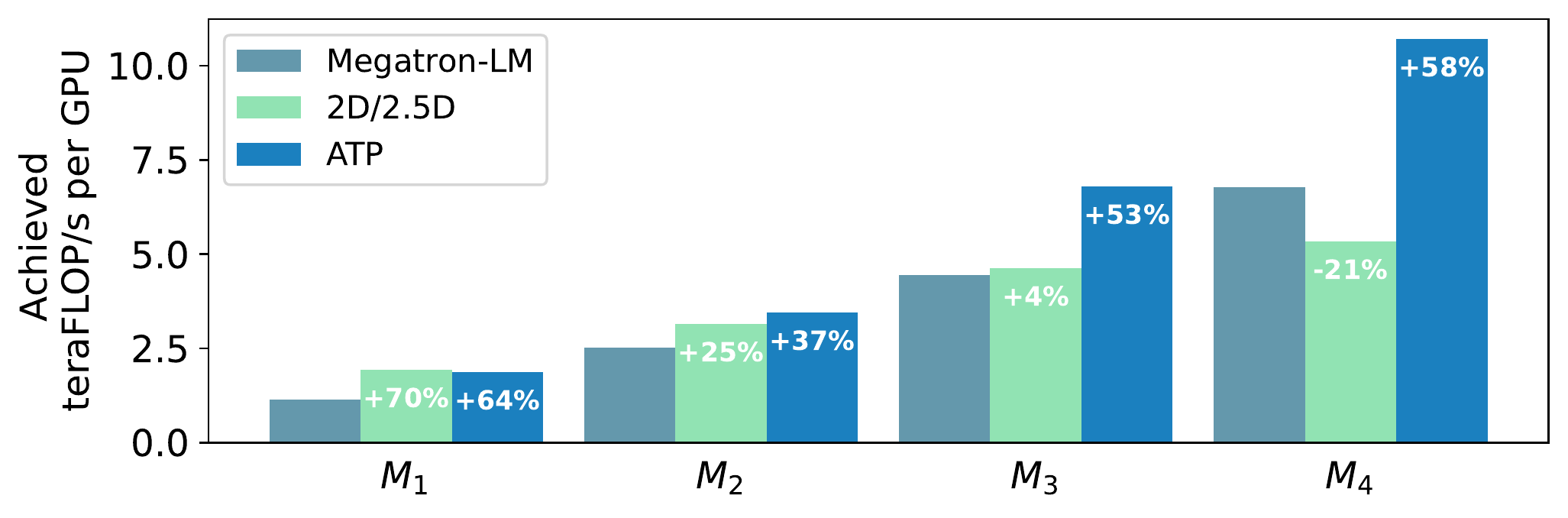}
  }
  \subfigure[$\bf IC_2$] {
    \label{fig:sota-t2}     
    \includegraphics[width=0.92\columnwidth]{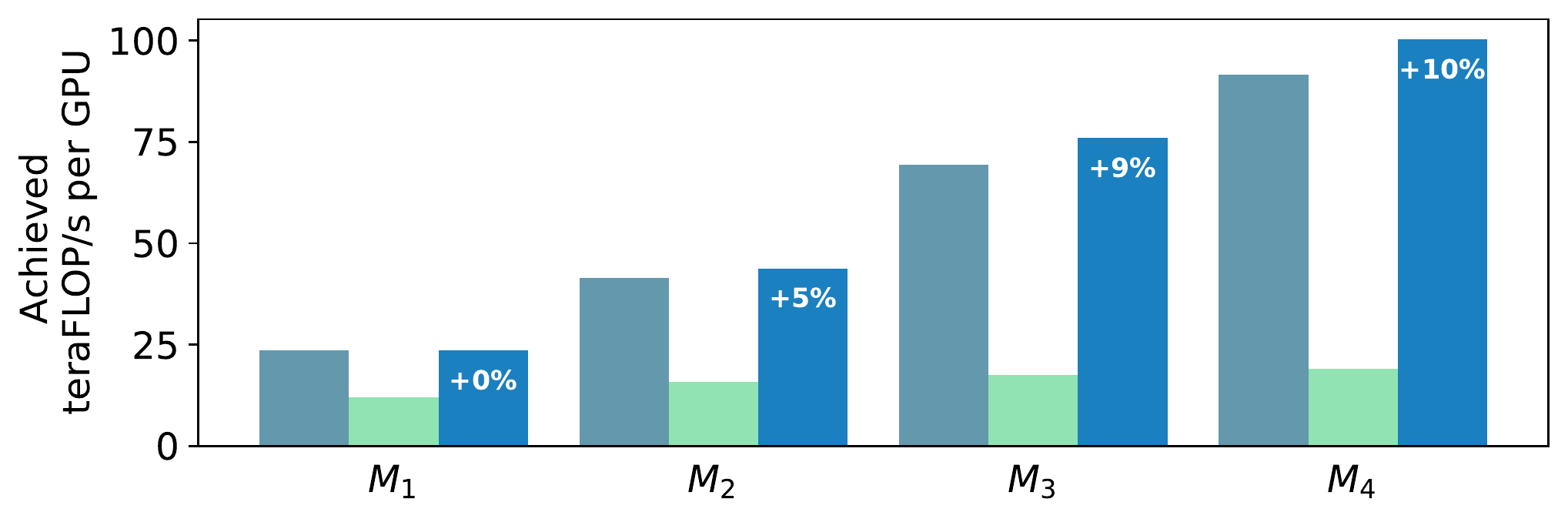}  
  }    
  \subfigure[$\bf IC_3$] { 
    \label{fig:sota-t3}     
    \includegraphics[width=0.92\columnwidth]{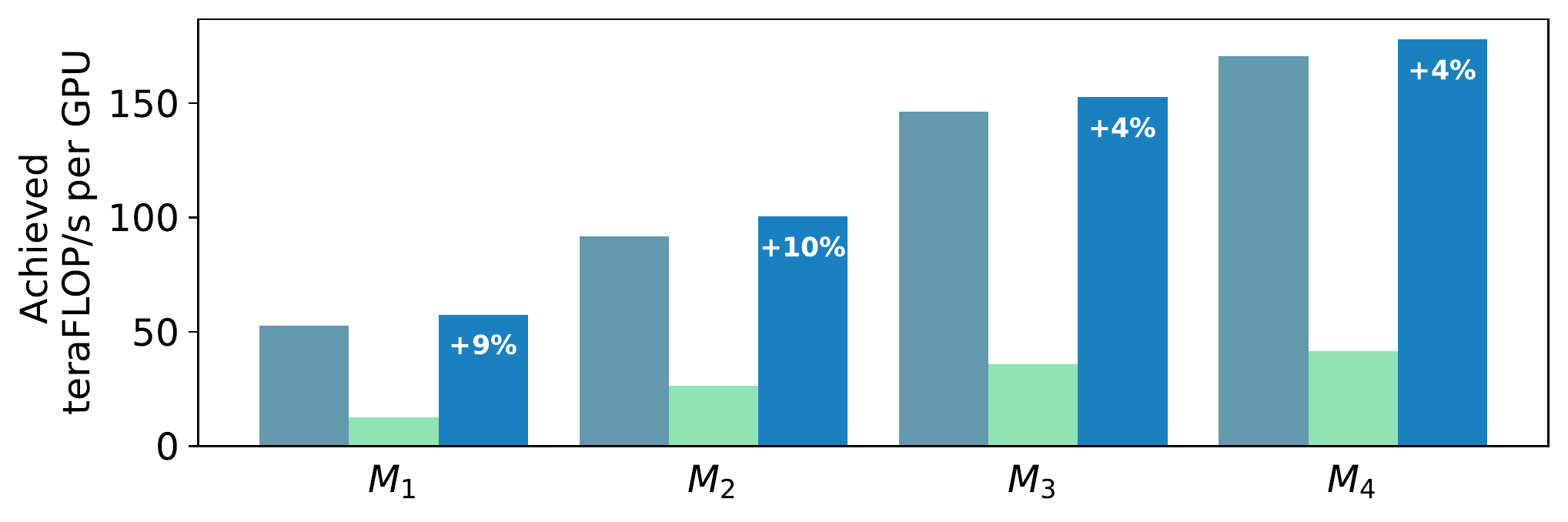}
  }
  \subfigure[$\bf IC_4$] {
    \label{fig:sota-t4}     
    \includegraphics[width=0.92\columnwidth]{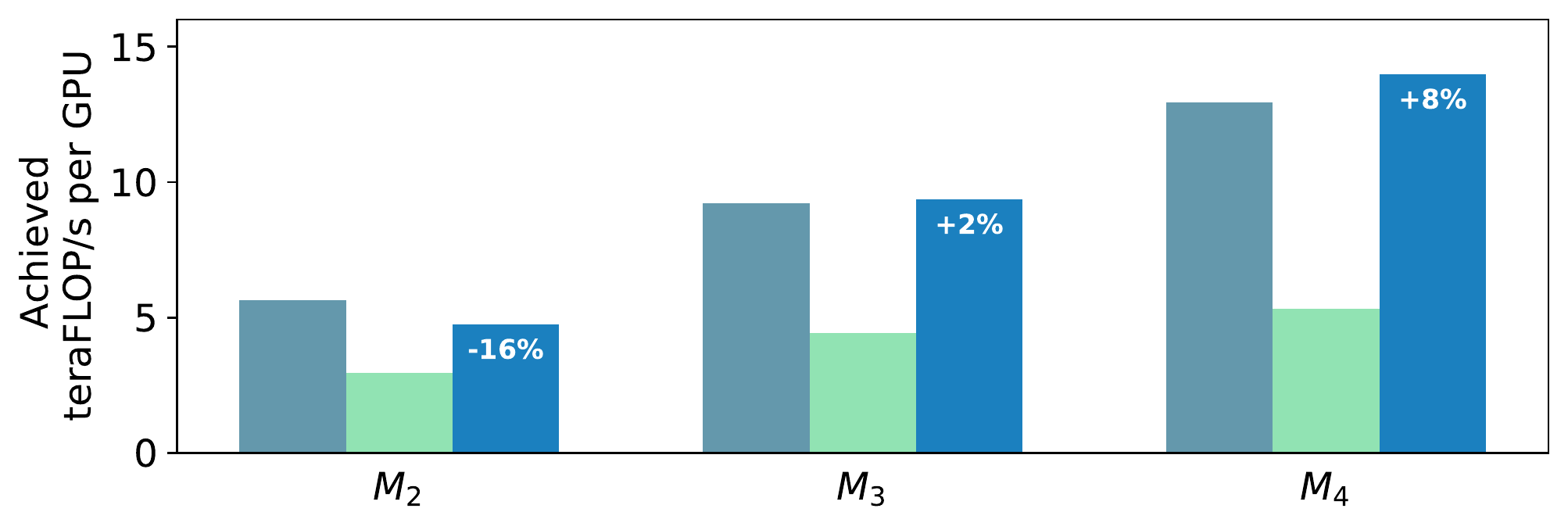}  
  }
  \vspace{-5pt}
  \caption{Comparison with the SOTA Methods. The percentages on the bars represent the performance improvement of ATP over Megatron-LM.} 
  \label{fig:sota}
\end{figure*}

To answer \textbf{Q1}, we compare ATP with Megatron-LM \cite{megatron} and 2D/2.5D tensor parallelism \cite{tp-2d, tp-25d}. In order to ensure fairness in the comparison, we do not use optimized kernels in any of the implementations. As baseline implementations, we use Megatron-LM public code v2.6 and used ColossalAI \cite{Bian2021ColossalAIAU} v0.1.12 for 2D/2.5D tensor parallelism. For each of the four interconnects ($IC_{1,2,3,4}$), we scale the training to the maximum number of GPUs: 8 GPUs for $IC_{1,2,3}$ and 16 GPUs for $IC_4$. We treat 2D/2.5D tensor parallelism as a single baseline, using 2D for 16 GPUs and 2.5D for 8 GPUs (2D tensor parallelism is only suitable for devices with square numbers).

As shown in Figure \ref{fig:sota}, ATP consistently outperform Megatron-LM and 2D/2.5D tensor parallelism across various topologies and model sizes. The optimization effect of ATP varies significantly depending on the interconnect topology. On PCIe-based $IC_1$, ATP demonstrated the greatest performance improvement, ranging from 37\% to 64\%. This is mainly due to the use of $DeviceMesh(2,4)$ in ATP, which significantly reduced communication cost. On NVLink-based $IC_{2,3}$, the performance improvement is smaller, with a maximum improvement of about 10\%. For these interconnects, ATP and Megatron-LM have similar communication costs because ATP select the optimal $DeviceMesh(N,1)$. The performance improvement in ATP is mainly due to better computation and communication overlap. On InfiniBand-based $IC_{4}$ (16 devices), ATP shows a 4\% improvement for larger models. A more detailed analysis can be found in Section \ref{5.3}.

2D/2.5D tensor parallelism performs significantly worse than Megatron-LM and ATP on $IC_{2,3,4}$ due to the reasons discussed in Section \ref{3.3}. On $IC_1$, ATP performs better than 2D/2.5D except for the smallest model size ($M_1$). As the model size increased, the performance advantage of 2D/2.5D relative to Megatron-LM decreases, and it performed poorly with $M_4$.

\subsection{Overlapping Optimization}

To investigate \textbf{Q2}, we compare ATP with different chunk sizes in Table \ref{tab:chunk_exp}. The table shows the training performance of $M_{2,3,4}$ on four interconnect topologies scaled to 8 GPUs. The \textit{baseline} represents the performance under the optimal ATP strategy, and then we add chunk optimization by setting the chunk size to 2 or 4.

\begin{table}[thb]
\small
\centering
\label{tab:chunk_exp}
\caption{Overlapping Optimization for Different Interconnect Topologies (TeraFLOP/s per GPU).}
\renewcommand\arraystretch{1.1}
\begin{tabular}{rlllll}
\noalign{\hrule height 1pt}
                       &         & $IC_1$    & $IC_2$ & $IC_3$ & $IC_4$ \\ \hline
\multirow{3}{*}{$M_2$}  & Baseline  & 3.45  & 42.82   & 97.45   & 15.31   \\
                       & Chunk=2  & 3.46  & 43.59   & \textbf{100.38}   & 17.20   \\
                       & Chunk=4  & \textbf{3.51}  & \textbf{43.73}   & 98.20   & \textbf{18.62}   \\ \hline
\multirow{3}{*}{$M_3$}  & Baseline  & 6.68  & 73.61   & 151.98   & 22.47   \\
                       & Chunk=2  & \textbf{6.79}  & 75.27   & \textbf{152.64}   & 26.09   \\
                       & Chunk=4  & 6.72  & \textbf{75.95}   & 151.79   & \textbf{26.18}   \\ \hline
\multirow{3}{*}{$M_4$} & Baseline  & \textbf{10.71} & 96.83   & \textbf{178.19}   & 32.52   \\
                       & Chunk=2  & 10.70 & \textbf{100.25}   & 178.17   & 34.95  \\
                       & Chunk=4  & 10.70 & 100.23   & 177.78   & \textbf{38.81}   \\ \noalign{\hrule height 1pt}
\end{tabular}
\end{table}

We make the following observations. \textbf{1)} The chunk-based overlapping optimization generally improves performance across different model sizes and interconnect topologies. \textbf{2)} the effect of chunk optimization is more significant on some interconnect, such as $IC_4$. On $IC_4$, larger chunk sizes also have greater performance improvements. For example, setting the chunk size to 4 results in a 16\% to 21\% improvement in end-to-end performance. \textbf{3)} For intra-node interconnects ($IC_{1,2,3}$), chunk-based overlapping typically resultes in a 1\% to 3\% improvement in end-to-end performance. \textbf{4)} In some cases, performance may degrade as the chunk size increases. This can be due to reduced parallelism, leading to inefficient hardware utilization for smaller GEMMs and other operators, as well as increased overhead at the framework level.

\begin{figure*}[thb]
  \centering
  \subfigure[$\bf IC_1$] { 
    \label{fig:mesh-t1}     
    \includegraphics[width=0.92\columnwidth]{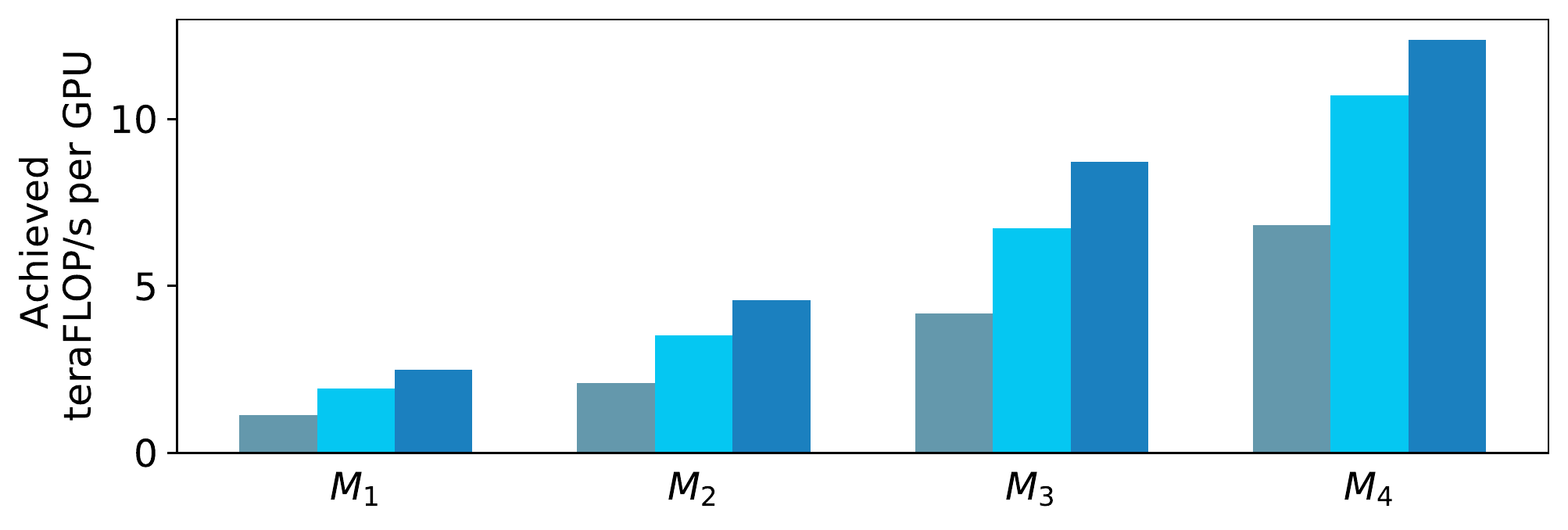}
  }
  \subfigure[$\bf IC_2$] {
    \label{fig:mesh-t2}     
    \includegraphics[width=0.92\columnwidth]{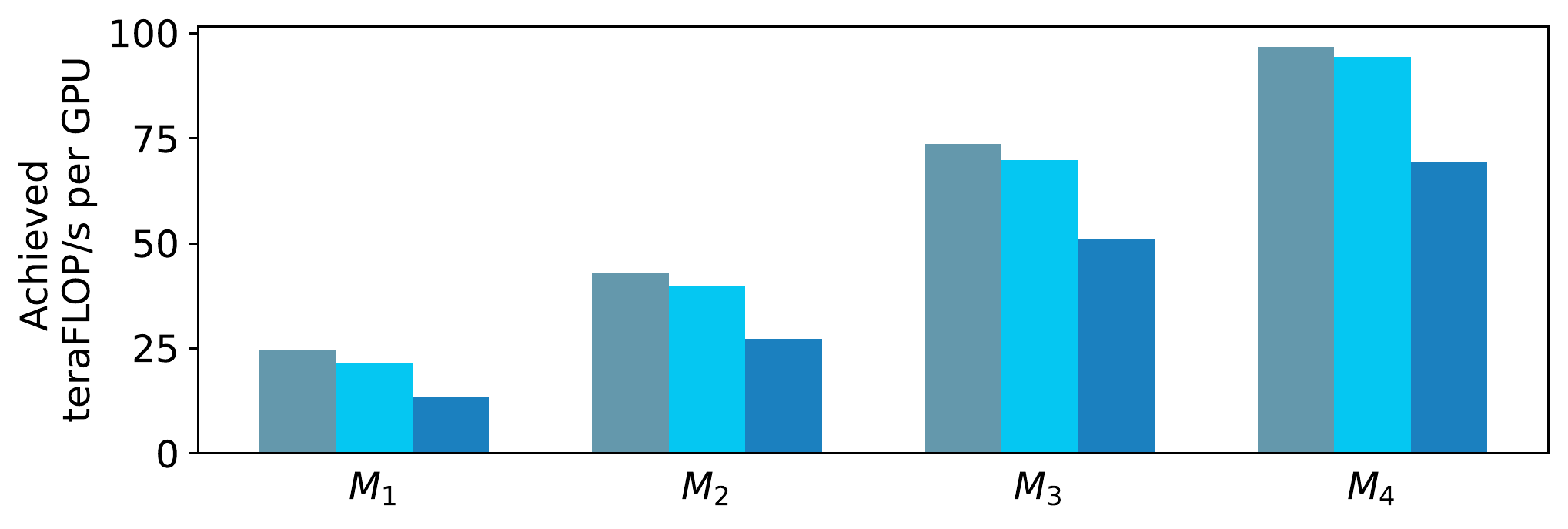}  
  }    
  \subfigure[$\bf IC_3$] { 
    \label{fig:mesh-t3}     
    \includegraphics[width=0.92\columnwidth]{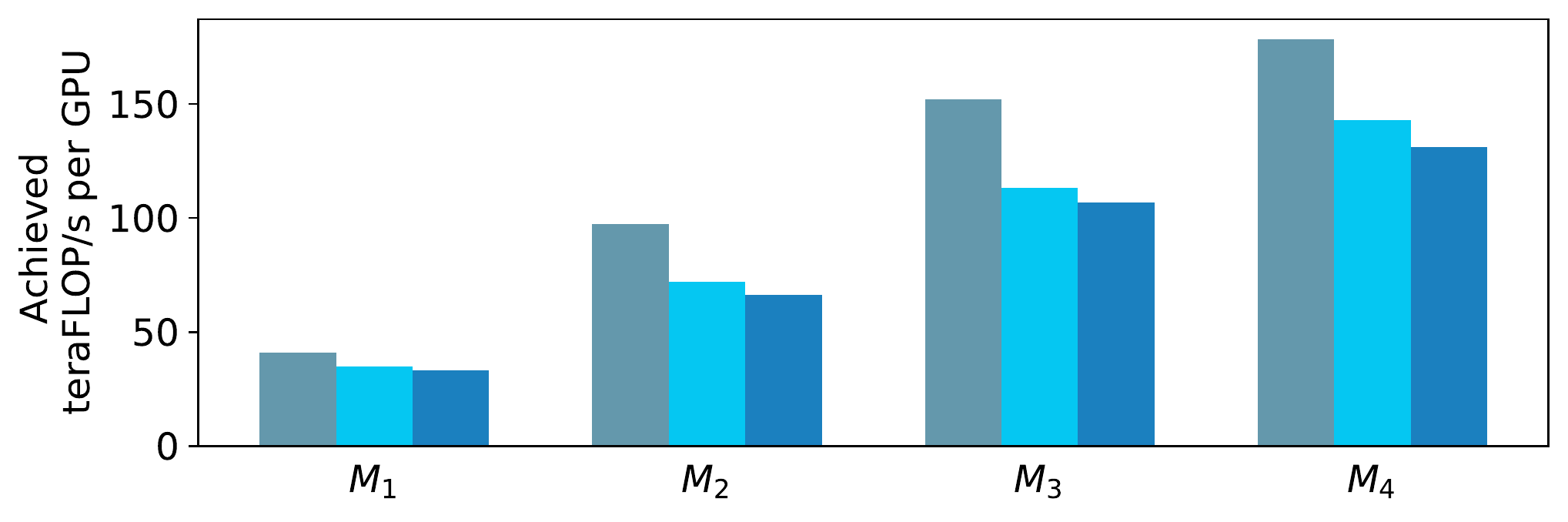}
  }
  \subfigure[$\bf IC_4$] {
    \label{fig:mesh-t4}     
    \includegraphics[width=0.92\columnwidth]{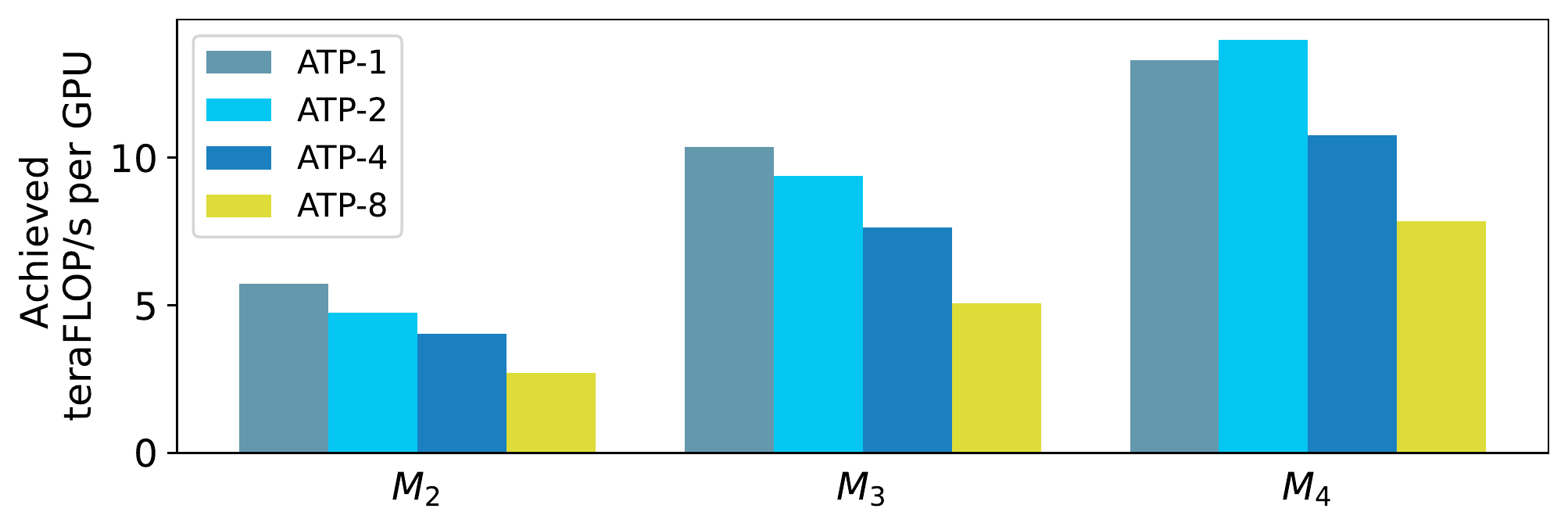}  
  }
  \caption{Performance comparison for ATP with different device meshes. ATP-$i$ refers to ATP with $DeviceMesh(N/i, i)$ ($N$ devices). For $IC_{1,2,3}$ with 8 devices, we compare ATP-1/2/4. For $IC_{4}$ with 16 devices, we compare ATP-1/2/4/8.} 
  \label{fig:mesh}
\end{figure*}

\subsection{Optimal Strategy for ATP}
\label{5.3}

To understand \textbf{Q3} and better compare the performance of ATP under different device mesh configurations, we evaluated the performance of different device meshes without communication optimization. The results for different interconnect topologies and model sizes are shown in Figure \ref{fig:mesh}. ATP-$i$ refers to ATP with $DeviceMesh(N/i,i)$ ($N$ devices).

For fully-connected topologies $IC_{3,4}$, the hierarchical communication matrix has only one layer. Using equations \ref{tcomm} and \ref{allreduce}, ATP selects ATP-1 when the number of devices is less than 8 and ATP-2 when it is greater than 8. The optimal ATP strategy is ATP-1 for $IC_3$ with 8 GPUs and ATP-2 for $IC_4$ with 16 GPUs. As shown in Figure \ref{fig:mesh-t3}, these results are consistent with the theoretical analysis. For $IC_4$, ATP-2 demonstrated a slight improvement on $M_4$. However, the performance degrades when the model is small due to the overhead introduced.

For the dual-GPUs interconnect $IC_2$, the hierarchical communication matrix has two layers, the first layer is 4 dual-GPUs interconnected by PCIe and the second layer consists of 2 GPUs connected by NVLink. We mainly compare the second item in formula \ref{tcomm}, $2h/d_2B_1$. Under ATP-1, this item is $2h/GroupBW_2$, while under ATP-2/4, we can get $B_1= GroupBW_2/d_2$ using formula \ref{bwd}, which is the same for all ATP strategies. However, the first item in ATP-1 is 0, making ATP-1 the most optimal strategy in this case. This trend is also shown in Figure \ref{fig:mesh-t2}.

For $IC_1$, we find that the all-reduce bandwidth could not be accurately evaluated using the hierarchical communication matrix. In this case, we need to calibrate the bandwidths $B_1$ and $B_2$ using measured performance. For $DeviceMesh(2,4)$, $B_1$ is calibrated to 1.20 GB/s and $B_2$ to 4.95 GB/s. For $DeviceMesh(8,1)$, $B_1$ is calibrated to 0.97 GB/s. Using Equation \ref{tcomm}, the $T_{comm}$ of ATP-4 was theoretically 46\% lower than that of ATP-1. As shown in Figure \ref{fig:mesh-t1}, the actual end-to-end training performance of ATP-4 was significantly improved, consistent with the calibrated theoretical analysis.

\subsection{Interconnect Topologies}

In the above experiments, all interconnects except $IC_1$ demonstrated relatively good performance under ATP-1. With the common GPU interconnect topologies currently available, ATP is optimal on $DeviceMesh(N,1)$ and does not show the advantage of ATP.

To investigate which interconnect topologies ATP can show superiority, we analyze the following two interconnect topologies from a theoretical perspective: NVLink-Network Switch ($\bf IC_5$) \cite{ishii2022nvlink} and 2D Torus ($\bf IC_6$). For $IC_5$, because all GPUs are interconnected through NVLink-Network Switch, they have the same communication bandwidth. Therefore, its hierarchical communication matrix has only one layer, and we can assume $B_1^{'}=B_2^{'}=GroupBW$. For $IC_6$, the hierarchical communication matrix is shown in \ref{fig:hcm-2}, in which case both $B_1^{'}$ and $B_2^{'}$ are also equal to $GroupBW$. Next, we combine equations \ref{tcomm} and \ref{allreduce} to obtain the tensor-parallel communication time for $DeviceMesh(d_1, d_2)$:

\[T_{comm} = \frac{2Lbsh}{GroupBW} \times \frac{14d_2+4d_1-18}{d_1d_2}\]

\begin{figure}[hbt]
    \centering
    \includegraphics[width=0.45\textwidth]{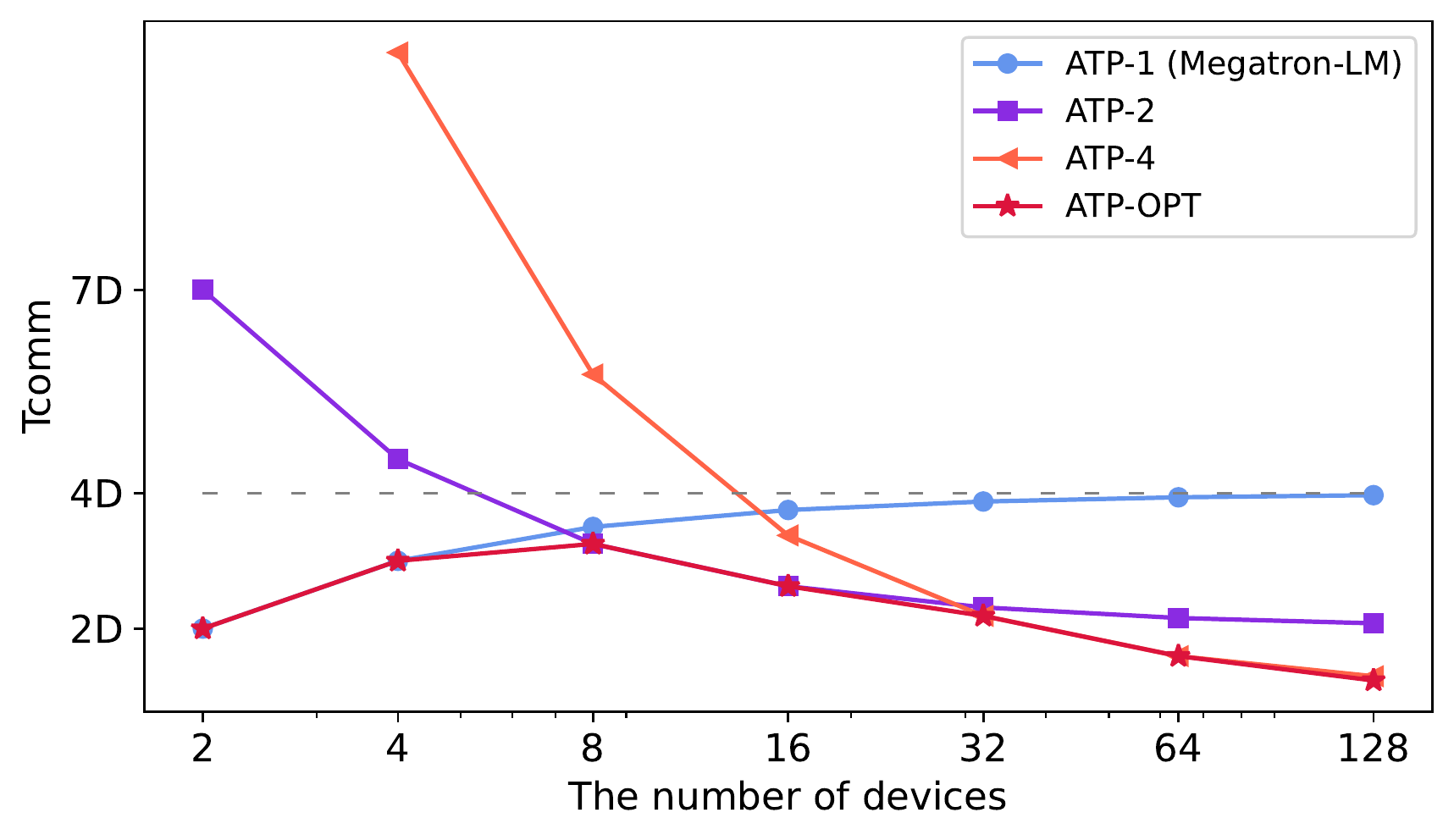}
    \vspace{-5pt}
    \caption{Communication time of $IC_{5,6}$ with increasing number of devices. ATP-$i$ refers to $DeviceMesh(N/i, i)$ and ATP-OPT refers to the optimal device mesh.}
    \label{fig:nvlink-switch}
\end{figure}

Assuming that $\Delta$ is equal to $2Lbsh / GroupBW$, we visualize the ATP under different device mesh in Figure \ref{fig:nvlink-switch}. We observe that the communication cost increases with scaling in ATP-1, while the cost decreases in ATP-2 and ATP-4. ATP-OPT corresponds to the communication overhead in the optimal device mesh. The communication cost of Megatron-LM has the same curve as ATP-1, with the communication cost rising with scaling. The communication cost of ATP gradually decreases as it scales up, and ATP will have a more significant advantage at larger scales.

In summary, in response to \textbf{Q4}, we suggest that future interconnect topologies should consider using accelerator-specific systems to avoid the bottlenecks of current CPU-oriented communication technologies. Additionally, fully-connected and torus topologies are expected to have good performance and scalability, with the latter being more cost-effective.

%% file: tex/6_conclusion.tex
\section{Related Work}

\textbf{Foundation Models.} Foundation models have been widely adopted in many fields such as language \cite{Devlin2019BERTPO, brown2020language, Wu2021Yuan1L}, vision \cite{Radford2021LearningTV, Yuan2021FlorenceAN}, and reasoning \cite{Polu2020GenerativeLM, Rabe2020MathematicalRV}. To train foundation models on ultra-large datasets and model sizes, custom software systems such as Megatron-LM \cite{megatron}, DeepSpeed \cite{Rasley2020DeepSpeedSO}, ColossalAI \cite{Bian2021ColossalAIAU} have been developed. These systems are built on parallelization techniques such as tensor parallelism \cite{megatron}, pipeline parallelism \cite{huang2019gpipe, narayanan2019pipedream} and ZeRO redundant optimizer \cite{Rajbhandari2019ZeROMO, Rajbhandari2021ZeROInfinityBT}. Tensor parallelism is critical for the efficiency of training giant foundation models, but it has been rarely studied systematically. ATP provides an adaptive tensor parallelism framework to improve training efficiency on complicated interconnection.

\textbf{Tensor Parallelism.} Megatron-LM introduced tensor parallelism for training GPT and BERT models, making it a standard technique for training foundation models. Several recent works, such as 2D tensor parallelism \cite{tp-2d} and 2.5D tensor parallelism \cite{tp-25d}, use the Scalable Universal Matrix Multiplication Algorithm (SUMMA) \cite{Geijn1995SUMMASU}. 2D tensor parallelism splits both activations and weights, and has a smaller memory footprint. 2.5D tensor parallelism extends the 2D approach to support an arbitrary number of GPUs, but still suffers from high communication overhead. Other works such as Sagemaker \cite{Karaku2021AmazonSM} and GSPMD \cite{Xu2021GSPMDGA} split activations along the hidden dimension, but also have inefficient communication. However, none of these approaches consider complicated GPU interconnects, leaving room for optimization. ATP addresses this by proposing a search space of tensor parallelism strategies and using a hierarchical communication matrix to estimate the communication cost of each strategy. ATP is more adaptable and outperforms other tensor parallelism methods in some environments with complicated GPU interconnects.

\textbf{Automatic Parallelism.} Recent work has proposed methods for optimizing distributed training through automatic parallelism. These methods, such as Alpa \cite{Zheng2022AlpaAI}, Unity \cite{Unger2022UnityAD}, GSPMD \cite{Xu2021GSPMDGA}, and Whale \cite{Jia2020WhaleEG}, can find the optimal hybrid parallelization strategy within a certain search space rather than relying on manually-designed strategies. However, these existing approaches only support regular tensor parallelism and simple performance model of communication. ATP offers a more general tensor Parallelism search space for automatic parallelism and can also better characterize the communication ability of complicated interconnects through the use of hierarchical communication matrix. This can improve the accuracy of these automatic parallelism methods in estimating communication cost.


\section{Conclusion}

In this work, we present Adaptive Tensor Parallelism (ATP) for foundation models. We propose column- and row-first tensor parallelism based on 2D device meshes, and use the hierarchical communication matrix to identify the optimal strategy with minimal communication cost on different interconnections. Additionally, we use chunk-based overlapping to reduce the communication overhead introduced by ATP. ATP outperforms state-of-the-art approaches for various interconnects and model sizes and shows theoretical improvement for some topologies. ATP can be implemented as a replacement for the current state-of-the-art tensor parallelism approach, providing improved performance, scalability, and adaptability to different (future) topologies.

%% file: main.bbl
\begin{thebibliography}{10}

\bibitem{Bian2021ColossalAIAU}
Zhengda Bian, Hongxin Liu, Boxiang Wang, Haichen Huang, Yongbin Li, Chuan-Qing
  Wang, Fan Cui, and Yang You.
\newblock Colossal-ai: A unified deep learning system for large-scale parallel
  training.
\newblock {\em ArXiv}, abs/2110.14883, 2021.

\bibitem{Bommasani2021OnTO}
Rishi Bommasani, Drew~A. Hudson, Ehsan Adeli, Russ Altman, Simran Arora, Sydney
  von Arx, Michael~S. Bernstein, Jeannette Bohg, Antoine Bosselut, Emma
  Brunskill, Erik Brynjolfsson, S.~Buch, Dallas Card, Rodrigo Castellon,
  Niladri~S. Chatterji, Annie~S. Chen, Kathleen~A. Creel, Jared Davis, Dora
  Demszky, Chris Donahue, Moussa Doumbouya, Esin Durmus, Stefano Ermon, John
  Etchemendy, Kawin Ethayarajh, Li~Fei-Fei, Chelsea Finn, Trevor Gale,
  Lauren~E. Gillespie, Karan Goel, Noah~D. Goodman, Shelby Grossman, Neel Guha,
  Tatsunori Hashimoto, Peter Henderson, John Hewitt, Daniel~E. Ho, Jenny Hong,
  Kyle Hsu, Jing Huang, Thomas~F. Icard, Saahil Jain, Dan Jurafsky, Pratyusha
  Kalluri, Siddharth Karamcheti, Geoff Keeling, Fereshte Khani, O.~Khattab,
  Pang~Wei Koh, Mark~S. Krass, Ranjay Krishna, Rohith Kuditipudi, Ananya Kumar,
  Faisal Ladhak, Mina Lee, Tony Lee, Jure Leskovec, Isabelle Levent, Xiang~Lisa
  Li, Xuechen Li, Tengyu Ma, Ali Malik, Christopher~D. Manning, Suvir~P.
  Mirchandani, Eric Mitchell, Zanele Munyikwa, Suraj Nair, Avanika Narayan,
  Deepak Narayanan, Benjamin Newman, Allen Nie, Juan~Carlos Niebles, Hamed
  Nilforoshan, J.~F. Nyarko, Giray Ogut, Laurel~J. Orr, Isabel Papadimitriou,
  Joon~Sung Park, Chris Piech, Eva Portelance, Christopher Potts, Aditi
  Raghunathan, Robert Reich, Hongyu Ren, Frieda Rong, Yusuf~H. Roohani, Camilo
  Ruiz, Jack Ryan, Christopher R'e, Dorsa Sadigh, Shiori Sagawa, Keshav
  Santhanam, Andy Shih, Krishna~Parasuram Srinivasan, Alex Tamkin, Rohan Taori,
  Armin~W. Thomas, Florian Tram{\`e}r, Rose~E. Wang, William Wang, Bohan Wu,
  Jiajun Wu, Yuhuai Wu, Sang~Michael Xie, Michihiro Yasunaga, Jiaxuan You,
  Matei~A. Zaharia, Michael Zhang, Tianyi Zhang, Xikun Zhang, Yuhui Zhang,
  Lucia Zheng, Kaitlyn Zhou, and Percy Liang.
\newblock On the opportunities and risks of foundation models.
\newblock {\em ArXiv}, abs/2108.07258, 2021.

\bibitem{brown2020language}
Tom Brown, Benjamin Mann, Nick Ryder, Melanie Subbiah, Jared~D Kaplan, Prafulla
  Dhariwal, Arvind Neelakantan, Pranav Shyam, Girish Sastry, Amanda Askell,
  et~al.
\newblock Language models are few-shot learners.
\newblock {\em Advances in neural information processing systems},
  33:1877--1901, 2020.

\bibitem{Chowdhery2022PaLMSL}
Aakanksha Chowdhery, Sharan Narang, Jacob Devlin, Maarten Bosma, Gaurav Mishra,
  Adam Roberts, Paul Barham, Hyung~Won Chung, Charles Sutton, Sebastian
  Gehrmann, Parker Schuh, Kensen Shi, Sasha Tsvyashchenko, Joshua Maynez,
  Abhishek~B Rao, Parker Barnes, Yi~Tay, Noam~M. Shazeer, Vinodkumar
  Prabhakaran, Emily Reif, Nan Du, Benton~C. Hutchinson, Reiner Pope, James
  Bradbury, Jacob Austin, Michael Isard, Guy Gur-Ari, Pengcheng Yin, Toju Duke,
  Anselm Levskaya, Sanjay Ghemawat, Sunipa Dev, Henryk Michalewski, Xavier
  Garc{\'i}a, Vedant Misra, Kevin Robinson, Liam Fedus, Denny Zhou, Daphne
  Ippolito, David Luan, Hyeontaek Lim, Barret Zoph, Alexander Spiridonov, Ryan
  Sepassi, David Dohan, Shivani Agrawal, Mark Omernick, Andrew~M. Dai,
  Thanumalayan~Sankaranarayana Pillai, Marie Pellat, Aitor Lewkowycz, Erica
  Moreira, Rewon Child, Oleksandr Polozov, Katherine Lee, Zongwei Zhou, Xuezhi
  Wang, Brennan Saeta, Mark D{\'i}az, Orhan Firat, Michele Catasta, Jason Wei,
  Kathleen~S. Meier-Hellstern, Douglas Eck, Jeff Dean, Slav Petrov, and Noah
  Fiedel.
\newblock Palm: Scaling language modeling with pathways.
\newblock {\em ArXiv}, abs/2204.02311, 2022.

\bibitem{Devlin2019BERTPO}
Jacob Devlin, Ming-Wei Chang, Kenton Lee, and Kristina Toutanova.
\newblock Bert: Pre-training of deep bidirectional transformers for language
  understanding.
\newblock {\em ArXiv}, abs/1810.04805, 2019.

\bibitem{10.5555/3571885.3571899}
Torsten Hoefler, Tommaso Bonato, Daniele De~Sensi, Salvatore Di~Girolamo,
  Shigang Li, Marco Heddes, Jon Belk, Deepak Goel, Miguel Castro, and Steve
  Scott.
\newblock Hammingmesh: A network topology for large-scale deep learning.
\newblock In {\em Proceedings of the International Conference on High
  Performance Computing, Networking, Storage and Analysis}, SC '22. IEEE Press,
  2022.

\bibitem{huang2019gpipe}
Yanping Huang, Youlong Cheng, Ankur Bapna, Orhan Firat, Dehao Chen, Mia Chen,
  HyoukJoong Lee, Jiquan Ngiam, Quoc~V Le, Yonghui Wu, et~al.
\newblock Gpipe: Efficient training of giant neural networks using pipeline
  parallelism.
\newblock {\em Advances in neural information processing systems}, 32, 2019.

\bibitem{InfiniBand2010IntroductionTI}
Paul~Grun InfiniBand.
\newblock Introduction to infiniband tm for end users industry-standard value
  and performance for high performance computing and the enterprise.
\newblock 2010.

\bibitem{ishii2022nvlink}
Alexander Ishii and Ryan Wells.
\newblock The nvlink-network switch: Nvidia’s switch chip for high
  communication-bandwidth superpods.
\newblock In {\em 2022 IEEE Hot Chips 34 Symposium (HCS)}, pages 1--23. IEEE
  Computer Society, 2022.

\bibitem{Jangda2021BreakingTC}
Abhinav Jangda, Jun Huang, Guodong Liu, Amir Hossein~Nodehi Sabet, Saeed
  Maleki, Youshan Miao, Madan Musuvathi, Todd Mytkowicz, and Olli Saarikivi.
\newblock Breaking the computation and communication abstraction barrier in
  distributed machine learning workloads.
\newblock {\em Proceedings of the 27th ACM International Conference on
  Architectural Support for Programming Languages and Operating Systems}, 2021.

\bibitem{Jia2020WhaleEG}
Xianyan Jia, Le~Jiang, Ang Wang, Wencong Xiao, Ziji Shi, J.~Zhang, Xinyuan Li,
  Lan yue Chen, Yong Li, Zhen Zheng, Xiaoyong Liu, and Wei Lin.
\newblock Whale: Efficient giant model training over heterogeneous gpus.
\newblock In {\em USENIX Annual Technical Conference}, 2020.

\bibitem{Jouppi2020ADS}
Norman~P. Jouppi, Doe~Hyun Yoon, George Kurian, Sheng Li, Nishant Patil, James
  Laudon, Cliff Young, and David~A. Patterson.
\newblock A domain-specific supercomputer for training deep neural networks.
\newblock {\em Communications of the ACM}, 63:67 -- 78, 2020.

\bibitem{Karaku2021AmazonSM}
Can~B{\"u}lent Karakuş, Rahul Huilgol, Fei Wu, Anirudh Subramanian, Cade
  Daniel, Derya Çavdar, Teng Xu, Haohan Chen, Arash Rahnama, and Luis~Carlos
  Quintela.
\newblock Amazon sagemaker model parallelism: A general and flexible framework
  for large model training.
\newblock {\em ArXiv}, abs/2111.05972, 2021.

\bibitem{narayanan2019pipedream}
Deepak Narayanan, Aaron Harlap, Amar Phanishayee, Vivek Seshadri, Nikhil~R
  Devanur, Gregory~R Ganger, Phillip~B Gibbons, and Matei Zaharia.
\newblock Pipedream: generalized pipeline parallelism for dnn training.
\newblock In {\em Proceedings of the 27th ACM Symposium on Operating Systems
  Principles}, pages 1--15, 2019.

\bibitem{megatron}
Deepak Narayanan, Mohammad Shoeybi, Jared Casper, Patrick LeGresley, Mostofa
  Patwary, Vijay Korthikanti, Dmitri Vainbrand, Prethvi Kashinkunti, Julie
  Bernauer, Bryan Catanzaro, et~al.
\newblock Efficient large-scale language model training on gpu clusters using
  megatron-lm.
\newblock In {\em Proceedings of the International Conference for High
  Performance Computing, Networking, Storage and Analysis}, pages 1--15, 2021.

\bibitem{nvswitch}
{NVIDIA Corporation}.
\newblock {\em NVIDIA NVSWITCH – The World’s Highest-Bandwidth On-Node
  Switch}.
\newblock NVIDIA Corporation.

\bibitem{Polu2020GenerativeLM}
Stanislas Polu and Ilya Sutskever.
\newblock Generative language modeling for automated theorem proving.
\newblock {\em ArXiv}, abs/2009.03393, 2020.

\bibitem{Pope2022EfficientlyST}
Reiner Pope, Sholto Douglas, Aakanksha Chowdhery, Jacob Devlin, James Bradbury,
  Anselm Levskaya, Jonathan Heek, Kefan Xiao, Shivani Agrawal, and Jeff Dean.
\newblock Efficiently scaling transformer inference.
\newblock {\em ArXiv}, abs/2211.05102, 2022.

\bibitem{Rabe2020MathematicalRV}
Markus~N. Rabe, Dennis Lee, Kshitij Bansal, and Christian Szegedy.
\newblock Mathematical reasoning via self-supervised skip-tree training.
\newblock {\em arXiv: Learning}, 2020.

\bibitem{Radford2021LearningTV}
Alec Radford, Jong~Wook Kim, Chris Hallacy, Aditya Ramesh, Gabriel Goh,
  Sandhini Agarwal, Girish Sastry, Amanda Askell, Pamela Mishkin, Jack Clark,
  Gretchen Krueger, and Ilya Sutskever.
\newblock Learning transferable visual models from natural language
  supervision.
\newblock In {\em International Conference on Machine Learning}, 2021.

\bibitem{Rajbhandari2019ZeROMO}
Samyam Rajbhandari, Jeff Rasley, Olatunji Ruwase, and Yuxiong He.
\newblock Zero: Memory optimizations toward training trillion parameter models.
\newblock {\em SC20: International Conference for High Performance Computing,
  Networking, Storage and Analysis}, pages 1--16, 2019.

\bibitem{Rajbhandari2021ZeROInfinityBT}
Samyam Rajbhandari, Olatunji Ruwase, Jeff Rasley, Shaden Smith, and Yuxiong He.
\newblock Zero-infinity: Breaking the gpu memory wall for extreme scale deep
  learning.
\newblock {\em SC21: International Conference for High Performance Computing,
  Networking, Storage and Analysis}, pages 1--15, 2021.

\bibitem{Rasley2020DeepSpeedSO}
Jeff Rasley, Samyam Rajbhandari, Olatunji Ruwase, and Yuxiong He.
\newblock Deepspeed: System optimizations enable training deep learning models
  with over 100 billion parameters.
\newblock {\em Proceedings of the 26th ACM SIGKDD International Conference on
  Knowledge Discovery \& Data Mining}, 2020.

\bibitem{Sergeev2018HorovodFA}
Alexander Sergeev and Mike~Del Balso.
\newblock Horovod: fast and easy distributed deep learning in tensorflow.
\newblock {\em ArXiv}, abs/1802.05799, 2018.

\bibitem{Shazeer2018MeshTensorFlowDL}
Noam~M. Shazeer, Youlong Cheng, Niki Parmar, Dustin Tran, Ashish Vaswani,
  Penporn Koanantakool, Peter Hawkins, HyoukJoong Lee, Mingsheng Hong, Cliff
  Young, Ryan Sepassi, and Blake~A. Hechtman.
\newblock Mesh-tensorflow: Deep learning for supercomputers.
\newblock {\em ArXiv}, abs/1811.02084, 2018.

\bibitem{Smith2022UsingDA}
Shaden Smith, Mostofa Patwary, Brandon Norick, Patrick LeGresley, Samyam
  Rajbhandari, Jared Casper, Zhun Liu, Shrimai Prabhumoye, George Zerveas,
  Vijay~Anand Korthikanti, Elton Zhang, Rewon Child, Reza~Yazdani Aminabadi,
  Julie Bernauer, Xia Song, Mohammad Shoeybi, Yuxiong He, Michael Houston,
  Saurabh Tiwary, and Bryan Catanzaro.
\newblock Using deepspeed and megatron to train megatron-turing nlg 530b, a
  large-scale generative language model.
\newblock {\em ArXiv}, abs/2201.11990, 2022.

\bibitem{Thakur2005OptimizationOC}
Rajeev Thakur, Rolf Rabenseifner, and William Gropp.
\newblock Optimization of collective communication operations in mpich.
\newblock {\em The International Journal of High Performance Computing
  Applications}, 19:49 -- 66, 2005.

\bibitem{Unger2022UnityAD}
Colin Unger, Zhihao Jia, Wei Wu, Sina Lin, Mandeep Baines, Carlos
  Efrain~Quintero Narvaez, Vinay~B. Ramakrishnaiah, Nirmal Prajapati,
  Patrick~S. McCormick, Jamaludin Mohd-Yusof, Xiaojian Luo, Jongsoo Park,
  Mikhail Smelyanskiy, and Alex Aiken.
\newblock Unity: Accelerating dnn training through joint optimization of
  algebraic transformations and parallelization.
\newblock In {\em USENIX Symposium on Operating Systems Design and
  Implementation}, 2022.

\bibitem{Geijn1995SUMMASU}
Robert~A. van~de Geijn and Jerrell Watts.
\newblock Summa: scalable universal matrix multiplication algorithm.
\newblock {\em Concurr. Pract. Exp.}, 9:255--274, 1995.

\bibitem{tp-25d}
Boxiang Wang, Qifan Xu, Zhengda Bian, and Yang You.
\newblock Tesseract: Parallelize the tensor parallelism efficiently.
\newblock In {\em {ICPP} 2022: 51th International Conference on Parallel
  Processing}, 2022.

\bibitem{Wu2021Yuan1L}
Shaohua Wu, Xudong Zhao, Tong Yu, Rongguo Zhang, Chong Shen, Hongli Liu, Feng
  Li, Hong Zhu, Jiangang Luo, Liang Xu, and Xuanwei Zhang.
\newblock Yuan 1.0: Large-scale pre-trained language model in zero-shot and
  few-shot learning.
\newblock {\em ArXiv}, abs/2110.04725, 2021.

\bibitem{tp-2d}
Qifan Xu, Shenggui Li, Chaoyu Gong, and Yang You.
\newblock An efficient 2d method for training super-large deep learning models.
\newblock {\em arXiv preprint arXiv:2104.05343}, 2021.

\bibitem{Xu2021GSPMDGA}
Yuanzhong Xu, HyoukJoong Lee, Dehao Chen, Blake~A. Hechtman, Yanping Huang,
  Rahul Joshi, Maxim Krikun, Dmitry Lepikhin, Andy Ly, Marcello Maggioni,
  Ruoming Pang, Noam~M. Shazeer, Shibo Wang, Tao Wang, Yonghui Wu, and Zhifeng
  Chen.
\newblock Gspmd: General and scalable parallelization for ml computation
  graphs.
\newblock {\em ArXiv}, abs/2105.04663, 2021.

\bibitem{Yuan2021OneFlowRT}
J.~Yuan, Xinqi Li, Cheng Cheng, Juncheng Liu, Ran Guo, Shenghang Cai, Chi Yao,
  Fei Yang, Xiaodong Yi, Chuan Wu, Haoran Zhang, and Jie Zhao.
\newblock Oneflow: Redesign the distributed deep learning framework from
  scratch.
\newblock {\em ArXiv}, abs/2110.15032, 2021.

\bibitem{Yuan2021FlorenceAN}
Lu~Yuan, Dongdong Chen, Yi-Ling Chen, Noel C.~F. Codella, Xiyang Dai, Jianfeng
  Gao, Houdong Hu, Xuedong Huang, Boxin Li, Chunyuan Li, Ce~Liu, Mengchen Liu,
  Zicheng Liu, Yumao Lu, Yu~Shi, Lijuan Wang, Jianfeng Wang, Bin Xiao, Zhen
  Xiao, Jianwei Yang, Michael Zeng, Luowei Zhou, and Pengchuan Zhang.
\newblock Florence: A new foundation model for computer vision.
\newblock {\em ArXiv}, abs/2111.11432, 2021.

\bibitem{Zhang2022OPTOP}
Susan Zhang, Stephen Roller, Naman Goyal, Mikel Artetxe, Moya Chen, Shuohui
  Chen, Christopher Dewan, Mona Diab, Xian Li, Xi~Victoria Lin, Todor Mihaylov,
  Myle Ott, Sam Shleifer, Kurt Shuster, Daniel Simig, Punit~Singh Koura, Anjali
  Sridhar, Tianlu Wang, and Luke Zettlemoyer.
\newblock Opt: Open pre-trained transformer language models.
\newblock {\em ArXiv}, abs/2205.01068, 2022.

\bibitem{Zheng2022AlpaAI}
Lianmin Zheng, Zhuohan Li, Hao Zhang, Yonghao Zhuang, Zhifeng Chen, Yanping
  Huang, Yida Wang, Yuanzhong Xu, Danyang Zhuo, Joseph Gonzalez, and Ion
  Stoica.
\newblock Alpa: Automating inter- and intra-operator parallelism for
  distributed deep learning.
\newblock In {\em USENIX Symposium on Operating Systems Design and
  Implementation}, 2022.

\bibitem{Ziakas2010IntelQI}
Dimitrios Ziakas, Allen~J. Baum, Robert~A. Maddox, and Robert~J. Safranek.
\newblock Intel{\textregistered} quickpath interconnect architectural features
  supporting scalable system architectures.
\newblock {\em 2010 18th IEEE Symposium on High Performance Interconnects},
  pages 1--6, 2010.

\end{thebibliography}
